\documentclass[fleqn]{article}
\usepackage[widespace]{fourier}
\usepackage[psamsfonts]{eucal}         
\usepackage{graphicx}
\usepackage{cite}
\usepackage[cmex10]{amsmath}
\usepackage{amssymb}
\usepackage{bm}
\usepackage{psfrag}
\usepackage{subfigure}
\usepackage[font=small]{caption}
\usepackage[usenames,dvipsnames]{color}
\renewcommand{\i}{\mathrm{i}}

\renewcommand{\vec}[1]{\boldsymbol{#1}}

\newcommand{\rmd}{d}

\newcommand{\x}{\boldsymbol{x}}
\newcommand{\q}{\boldsymbol{q}}

\newcommand{\kbf}{\boldsymbol{k}}
\newcommand{\psihat}{\hat{\psi}}
\newcommand{\phihat}{\hat{\phi}}
\newcommand{\ofx}{(\x)}
\newcommand{\etal}{\emph{et al}}

\newcommand\BECn{Bose--Einstein condensation}

\newcommand{\paperi}{\emph{Paper I}}
\newcommand{\paperiii}{\emph{Paper III}}
\newcommand{\fig}[1]{Fig.\,{#1}}

\title{Effective Field Theory for Atom-Molecule Systems II: Stationary Solutions and Bogoliubov
Excitations in Atom-Molecule Systems} 
\author{Catarina E Sahlberg and C W Gardiner
\\[5mm] \emph{Jack Dodd Centre for Quantum Technology,} 
\\ \emph{Department of Physics, University of Otago,}
\\ \emph{ Dunedin, New Zealand}}
\begin{document}
\maketitle

\begin{abstract}
We formulate the basic theoretical methods for \BECn\ of atoms 
close to a Feshbach resonance, in which the tunable scattering length
of the atoms is described using a system of coupled atom and molecule
fields.  These include the Thomas--Fermi description of the condensate
profile, the c-field equations, and the Bogoliubov--de Gennes
equations, and the Bogoliubov excitation spectrum for a homogenous
condensed system.   We apply this formalism to the special case of
Bragg scattering from a uniform condensate, and find that for moderate
and large scattering lengths, there is a dramatic difference in the
shift of the peak of the Bragg spectra, compared to that based on a
structureless atom model.  The result is compatible with the
experimental results of \cite{papp2008}.
\end{abstract}

\section{Introduction}
A recent experiment on Bragg scattering of a Bose-Einstein condensate
(BEC) close to a Feshbach resonance \cite{papp2008}, investigated the
effect of varying the interatomic scattering length on the
position of the Bragg spectral peak. This experiment used values of
scattering length, much larger than have previously been used in
Bragg scattering experiments, and noted that there was no
theoretically justifiable method of describing a BEC in this regime.

In \paperi\ \cite{sahlberg1} we formulated a method for treating ultracold
atoms close to a Feshbach resonance as a system of coupled atom and
molecule fields.  This method is applicable in the regime covered by
the experiment, and has been developed in such a a way as to be
directly applicable to a c-field description of a BEC in the vicinity
of a Feshbach resonance. The aim of this paper is to formulate the
standard results, such as the Thomas--Fermi profiles and the Bogoliubov spectrum, appropriate to this
model of systems of condensed atoms and molecules, and describe the differences
which arise compared to the corresponding theories of structureless
atoms with the same scattering length.

This paper is a precursor to our next paper, in which full c-field
computations of Bragg scattering will be performed, in a way which is
directly comparable with the experiment of \cite{papp2008}. Here, we
apply this model to the case of ideal Bragg scattering from a
uniform condensate, and compare this with the corresponding
theory for a BEC of structureless atoms, as done previously in
\cite{blakie2002}.  By doing this we obtain a benchmark comparison of
the behaviour of a Feschbach molecule treatment with that of the
structureless atom models, avoiding the numerous complexities which
necessarily arise in experiments. Thus, we can isolate the effects
which arise \emph{only} from the dynamics of the molecular basis of
the Feshbach resonance, from those which are forced on us by the
practicalities of experimental procedure.

\section{Formalism}
\label{sec:formalism}%
The formalism used here is outlined in \paperi\ \cite{sahlberg1} and is based
on the c-field methods described in detail in \cite{blakie2008}.

\subsection{Phenomenological Hamiltonian for atom-molecule system}
As introduced in \paperi, the Hamiltonian of a coupled atom-molecule
system is given by
\begin{eqnarray}
\label{eq:H}
\hat{H} = \int{\rm\rmd\x\,\,
\left\{\psihat^\dagger\ofx\left(-\frac{\hbar^2\nabla^2}{2m}+V_a\ofx\right)
\psihat\ofx
+\phihat^\dagger\ofx\left(-\frac{\hbar^2\nabla^2}{4m}+V_m\ofx+\varepsilon
\right)
\phihat\ofx
+ \right.} \nonumber \\
\left.\frac{U_{aa}}{2}\psihat^\dagger\ofx\psihat^\dagger\ofx\psihat\ofx\psihat
\ofx
+
{U_{am}}\psihat^\dagger\ofx\phihat^\dagger\ofx\phihat\ofx\psihat\ofx
+ \right. \nonumber \\
\left.\frac{U_{mm}}{2}\phihat^\dagger\ofx\phihat^\dagger\ofx\phihat\ofx\phihat
\ofx
+ \frac{g}{2}\left(\phihat^\dagger\ofx\psihat\ofx\psihat\ofx +
\psihat^\dagger\ofx\psihat^\dagger\ofx\phihat\ofx\right) \right\},
\end{eqnarray}
where $U_{aa}=4\pi\hbar^2a_{bg}/m$ is the background interaction
strength and $V_a$ and $V_m$ are the external trapping potential for
the atoms and molecules respectively.  The terms with factors $U_{am}$
and $U_{mm}$ correspond to atom-molecule and molecule-molecule
scattering. However, in the systems considered here, the molecule
field arises only during
collisions, and is very small, making these terms negligible.

\subsubsection{Values of parameters}
The parameters $g$ and $\varepsilon$ are the coupling strength and
detuning in the Feshbach resonance respectively.  In our formalism
they are given in terms of the experimentally measurable binding
energy
and s-wave scattering length by
\begin{eqnarray}
\label{eq:epsilon}
\varepsilon &=&{\hbar^2\alpha^2\over 2m}
\frac{\left(\pi-2\Lambda a_s\right)
\left(1-{2\Lambda a_{bg}} t(\frac{\alpha}{\Lambda})/{\pi}\right)}
{\Lambda a_s(1+t(\frac{\alpha}{\Lambda}))-\pi} 
\,,\\
\label{eq:g2}
g^2
&=& {8\pi\hbar^4\alpha^2\over m^2}
{\left(a_{bg}(\pi-2\Lambda a_s) -\pi a_s\right)
\left(1-{2\Lambda a_{bg}} t(\frac{\alpha}{\Lambda})/{\pi}\right)
\over
2\Lambda a_s \left(1+t(\frac{\alpha}{\Lambda})\right) -\pi}\, .
\end{eqnarray}
where $t(x)=x-\arctan{1/x}$, the parameter $\Lambda$ is the momentum
space cutoff, and $\hbar^2\alpha^2/m$ is the molecule binding energy
for a certain value of the s-wave scattering length $a_s$
\cite{sahlberg1}.  It is important to note that in this
phenomenological model the parameters $\varepsilon$ and $g$ take on
different values, depending on the closeness to the Feshbach
resonance.

\begin{figure}[t]
\begin{center}
\subfigure[]{
\begin{psfrags}%
\psfrag{s01}[t][t]{\color[rgb]{0,0,0}\setlength{\tabcolsep}{0pt}\begin{tabular}{c}Inverse scattering length $1/a_s$ $[a_0^{-1}]$\end{tabular}}%
\psfrag{s02}[b][b]{\color[rgb]{0,0,0}\setlength{\tabcolsep}{0pt}\begin{tabular}{c}Detuning $\varepsilon$ [MHz]\end{tabular}}%
\psfrag{x01}[t][t]{0}%
\psfrag{x02}[t][t]{0.2}%
\psfrag{x03}[t][t]{0.4}%
\psfrag{x04}[t][t]{0.6}%
\psfrag{x05}[t][t]{0.8}%
\psfrag{x06}[t][t]{1}%
\psfrag{x07}[t][t]{0}%
\psfrag{x08}[t][t]{0.005}%
\psfrag{x09}[t][t]{0.01}%
\psfrag{v01}[r][r]{0}%
\psfrag{v02}[r][r]{0.2}%
\psfrag{v03}[r][r]{0.4}%
\psfrag{v04}[r][r]{0.6}%
\psfrag{v05}[r][r]{0.8}%
\psfrag{v06}[r][r]{1}%
\psfrag{v07}[r][r]{-10}%
\psfrag{v08}[r][r]{-5}%
\psfrag{v09}[r][r]{0}%
\resizebox{6cm}{!}{\includegraphics{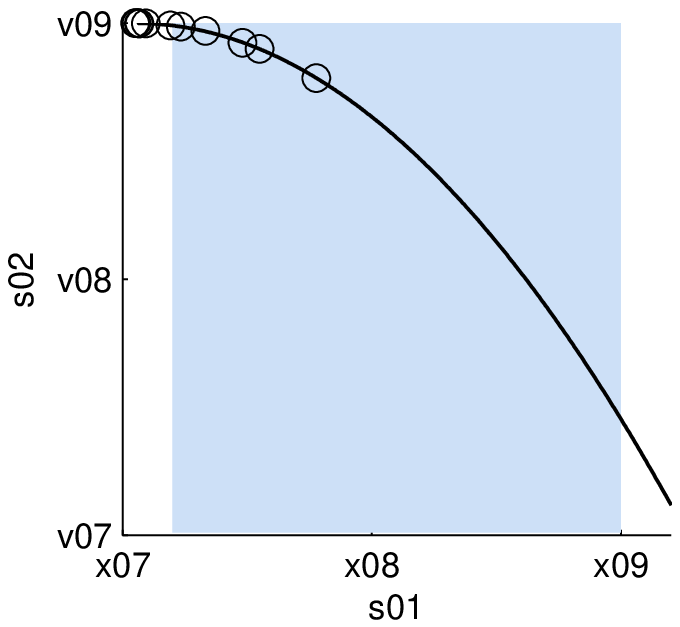}}%
\end{psfrags}%
\label{fig:epsilon}}\hspace{0.2in}
\subfigure[]{
\begin{psfrags}%
\psfrag{s01}[t][t]{\color[rgb]{0,0,0}\setlength{\tabcolsep}{0pt}\begin{tabular}{c}Inverse scattering length $1/a_s$ $[a_0^{-1}]$\end{tabular}}%
\psfrag{s02}[b][b]{\color[rgb]{0,0,0}\setlength{\tabcolsep}{0pt}\begin{tabular}{c}Coupling $g$ [Jm$^{-3/2}$]\end{tabular}}%
\psfrag{s09}[lt][lt]{\color[rgb]{0,0,0}\setlength{\tabcolsep}{0pt}\begin{tabular}{l}$10^{-38}$\end{tabular}}%
\psfrag{x01}[t][t]{0}%
\psfrag{x02}[t][t]{0.2}%
\psfrag{x03}[t][t]{0.4}%
\psfrag{x04}[t][t]{0.6}%
\psfrag{x05}[t][t]{0.8}%
\psfrag{x06}[t][t]{1}%
\psfrag{x07}[t][t]{0}%
\psfrag{x08}[t][t]{0.005}%
\psfrag{x09}[t][t]{0.01}%
\psfrag{v01}[r][r]{0}%
\psfrag{v02}[r][r]{0.2}%
\psfrag{v03}[r][r]{0.4}%
\psfrag{v04}[r][r]{0.6}%
\psfrag{v05}[r][r]{0.8}%
\psfrag{v06}[r][r]{1}%
\psfrag{v07}[r][r]{0}%
\psfrag{v08}[r][r]{0.5}%
\psfrag{v09}[r][r]{1}%
\psfrag{v10}[r][r]{1.5}%
\psfrag{v11}[r][r]{2}%
\psfrag{v12}[r][r]{2.5}%
\resizebox{6cm}{!}{\includegraphics{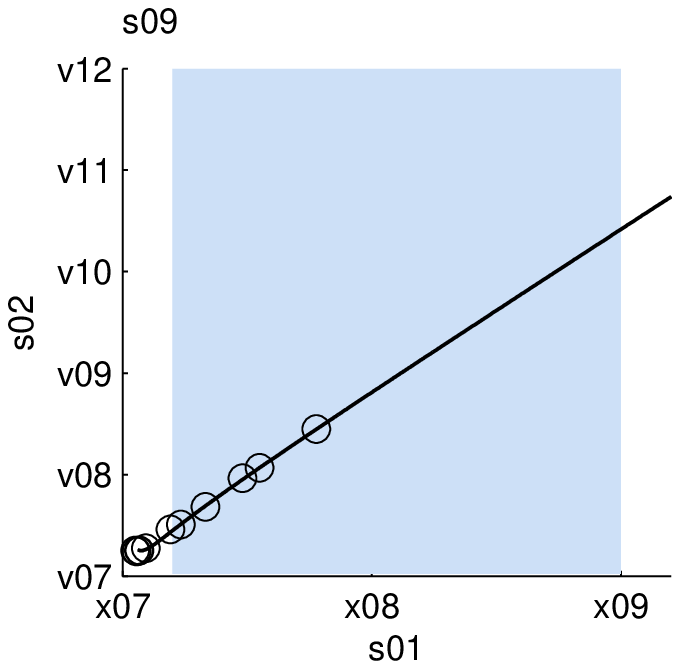}}%
\end{psfrags}%
\label{fig:g}}
\caption{The Feshbach resonance detuning $\varepsilon$, and coupling
parameter $g$, as functions of the inverse scattering length. The
black circles indicate the values calculated using the experimental
data from \cite{claussen2003} and equations (\ref{eq:g2}) and
(\ref{eq:epsilon}). The black lines are fits to these calculated
values. The shaded blue areas indicate the range of scattering
lengths of interest in the experiment of \cite{papp2008}, as well as
in the later sections of this paper. The momentum cutoff is here
chosen to be $\Lambda = 10^6$ m$^{-1}$.}
\label{fig:Parameters}
\end{center}
\end{figure}

In this paper we will consider a BEC of $^{85}$Rb, for which there is
experimental data of the binding energy close to the Feshbach
resonance at 155 G \cite{claussen2003}.  For a given momentum cutoff
$\Lambda$ we can then calculate the values of the coupling $g$, and
detuning $\varepsilon$, for each value of the scattering length, using
equations (\ref{eq:g2}) and (\ref{eq:epsilon}). We fit curves to
the values calculated from the experimental data in order to
extrapolate to other values of the scattering length.  The data,
obtained using (\ref{eq:epsilon}) and (\ref{eq:g2}) are well fitted by a
linear relationship between $g$ and $a_{s}^{-1}$, and a quadratic
relationship between $\varepsilon$ and $a_{s}^{-1}$. We show the
data and fits in \fig{\ref{fig:Parameters}}.

\subsubsection{C-field equations}
Since the atom field is usually much
larger than the molecule field in the situations we shall consider,
we can set $U_{am}$ and $U_{mm}$ equal to zero and the c-field
equations of motion corresponding to the
Hamiltonian (\ref{eq:H}) become
\begin{eqnarray}
\label{eq:EqsMotPsi}
i\hbar\frac{\partial{\psi}(\x)}{\partial t} &=&
\left(-\frac{\hbar^{2}\nabla^{2}}{2m}+V_{a}(\x)\right){\psi}(\x)+
U_{aa}|{\psi}(\x)|^{2}{\psi}(\x)+g{\psi}(\x)^{*}{\phi}(\x), \\
\label{eq:EqsMotPhi}
i\hbar\frac{\partial{\phi}(\x)}{\partial t} &=&
\left(-\frac{\hbar^{2}\nabla^{2}}{4m}+\varepsilon+V_{m}(\x)\right){\phi}(\x)+
\frac{g}{2}{\psi}^2(\x).
\end{eqnarray}
In practice, when solving these equations numerically, it is
necessary to explicitly use a projector in order to restrict the
wavefunctions to
the momentum subspace below the cutoff $\Lambda$, as described in
\paperi\ and implemented in \paperiii. 

The value of the momentum space cutoff $\Lambda$ will, in the case of
numerical simulation, arise from the simulation grid; any such
computation is restricted to a finite number of momentum space modes. It is also necessary to introduce a projector in order to avoid
effects from aliasing \cite{blakie2008}.  Furthermore, a momentum
cutoff is necessary in order for the pseudopotential approximation to
be valid \cite{braaten2006}.  However, as we show in \paperi, the
actual choice of value for the cutoff required for our
simulations has only a very small effect on the values of the
phenomenological Hamiltonian parameters.

\subsection{Thomas-Fermi Solutions}
\label{sec:ThomasFermi}%
Stationary solutions to the equations of motion (\ref{eq:EqsMotPsi})
and (\ref{eq:EqsMotPhi}) can be obtained by letting the time evolution
of the wavefunctions be $\psi(\x,t)=\psi\ofx \exp(i\mu_at/\hbar)$ and
$\phi(\x,t)=\phi\ofx \exp(i\mu_mt/\hbar)$.  It is clear that $\mu_m =
2\mu_a$ in order for the coupling terms to be time-independent, so we
get
\begin{eqnarray}
\mu_{a}\psi\ofx &=&
\left(-\frac{\hbar^{2}\nabla^{2}}{2m}+V_{a}(\x)\right)\psi\ofx+U_{aa}|\psi\ofx|^{2}\psi\ofx
+g\psi\ofx^{*}\phi\ofx, \\
2\mu_{a}\phi\ofx &=&
\left(-\frac{\hbar^{2}\nabla^{2}}{4m}+\varepsilon+V_{m}(\x)\right)\phi\ofx
+\frac{g}{2}\psi\ofx^{2}.
\end{eqnarray}
Taking the Thomas-Fermi limit and solving for $\psi$ and $\phi$ gives
the Thomas-Fermi solutions
\begin{eqnarray}
\label{eq:TF_psi}
|\psi_s\ofx| &=& \sqrt{\frac{\mu_a - V_{a}\ofx}{U_{aa}
+g^2/2(2\mu_a-\varepsilon - V_{m}\ofx)}}, \\
\label{eq:TF_phi}
\phi_s\ofx &=& \frac{g}{2(2\mu_a-\varepsilon - V_{m}\ofx)}\psi^2\ofx.
\end{eqnarray}
\begin{figure}[t]
\begin{center}
\begin{psfrags}%
\psfrag{s01}[b][b]{\color[rgb]{0,0,0}\setlength{\tabcolsep}{0pt}\begin{tabular}{c}Wavefunctions $[x_0^{-3/2}]$\end{tabular}}%
\psfrag{s02}[t][t]{\color[rgb]{0,0,0}\setlength{\tabcolsep}{0pt}\begin{tabular}{c}Radius $r$ $[x_0]$\end{tabular}}%
\psfrag{s03}[b][b]{\color[rgb]{0,0,0}\setlength{\tabcolsep}{0pt}\begin{tabular}{c}Scattering length $855a_0$\end{tabular}}%
\psfrag{s05}[b][b]{\color[rgb]{0,0,0}\setlength{\tabcolsep}{0pt}\begin{tabular}{c}Wavefunctions $[x_0^{-3/2}]$\end{tabular}}%
\psfrag{s06}[t][t]{\color[rgb]{0,0,0}\setlength{\tabcolsep}{0pt}\begin{tabular}{c}Radius $r$ $[x_0]$\end{tabular}}%
\psfrag{s07}[b][b]{\color[rgb]{0,0,0}\setlength{\tabcolsep}{0pt}\begin{tabular}{c}Scattering length $2133a_0$\end{tabular}}%
\psfrag{s17}[lt][lt]{\color[rgb]{0,0,0}\setlength{\tabcolsep}{0pt}\begin{tabular}{l}$\phi$\end{tabular}}%
\psfrag{s18}[lt][lt]{\color[rgb]{0,0,0}\setlength{\tabcolsep}{0pt}\begin{tabular}{l}$\psi_{GPE}$\end{tabular}}%
\psfrag{s19}[lt][lt]{\color[rgb]{0,0,0}\setlength{\tabcolsep}{0pt}\begin{tabular}{l}$\psi$\end{tabular}}%
\psfrag{s20}[lt][lt]{\color[rgb]{0,0,0}\setlength{\tabcolsep}{0pt}\begin{tabular}{l}$\phi$\end{tabular}}%
\psfrag{s25}[lt][lt]{\color[rgb]{0,0,0}\setlength{\tabcolsep}{0pt}\begin{tabular}{l}$\psi_{GPE}$\end{tabular}}%
\psfrag{s26}[lt][lt]{\color[rgb]{0,0,0}\setlength{\tabcolsep}{0pt}\begin{tabular}{l}$\psi$\end{tabular}}%
\psfrag{x01}[t][t]{0}%
\psfrag{x02}[t][t]{0.1}%
\psfrag{x03}[t][t]{0.2}%
\psfrag{x04}[t][t]{0.3}%
\psfrag{x05}[t][t]{0.4}%
\psfrag{x06}[t][t]{0.5}%
\psfrag{x07}[t][t]{0.6}%
\psfrag{x08}[t][t]{0.7}%
\psfrag{x09}[t][t]{0.8}%
\psfrag{x10}[t][t]{0.9}%
\psfrag{x11}[t][t]{1}%
\psfrag{x12}[t][t]{0}%
\psfrag{x13}[t][t]{5}%
\psfrag{x14}[t][t]{10}%
\psfrag{x15}[t][t]{15}%
\psfrag{x16}[t][t]{0}%
\psfrag{x17}[t][t]{5}%
\psfrag{x18}[t][t]{10}%
\psfrag{x19}[t][t]{15}%
\psfrag{v01}[r][r]{0}%
\psfrag{v02}[r][r]{0.1}%
\psfrag{v03}[r][r]{0.2}%
\psfrag{v04}[r][r]{0.3}%
\psfrag{v05}[r][r]{0.4}%
\psfrag{v06}[r][r]{0.5}%
\psfrag{v07}[r][r]{0.6}%
\psfrag{v08}[r][r]{0.7}%
\psfrag{v09}[r][r]{0.8}%
\psfrag{v10}[r][r]{0.9}%
\psfrag{v11}[r][r]{1}%
\psfrag{v12}[r][r]{0}%
\psfrag{v13}[r][r]{1}%
\psfrag{v14}[r][r]{2}%
\psfrag{v15}[r][r]{3}%
\psfrag{v16}[r][r]{4}%
\psfrag{v17}[r][r]{5}%
\psfrag{v18}[r][r]{6}%
\psfrag{v19}[r][r]{7}%
\psfrag{v20}[r][r]{0}%
\psfrag{v21}[r][r]{1}%
\psfrag{v22}[r][r]{2}%
\psfrag{v23}[r][r]{3}%
\psfrag{v24}[r][r]{4}%
\psfrag{v25}[r][r]{5}%
\psfrag{v26}[r][r]{6}%
\psfrag{v27}[r][r]{7}%
\resizebox{12cm}{!}{\includegraphics{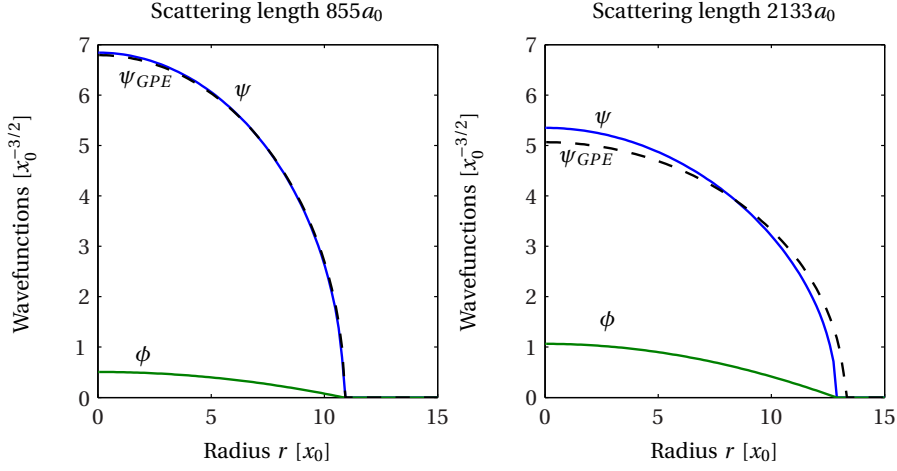}}%
\end{psfrags}%
\caption{Thomas-Fermi profiles for a
spherically symmetric condensate of $^{85}$Rb for two of the different
scattering lengths reported in Ref.\,\cite{claussen2003}: $a_s = 856a_0$ (left panel) and $a_s = 2133a_0$
(right panel).  The trapping frequency is $\omega_x=2\pi\times17.5$
Hz, and the total particle number --- counting each molecule as two
atoms --- is $10^5$.  The blue solid lines are the atom Thomas-Fermi
profiles from (\ref{eq:TF_psi}), the green solid lines are the
molecule profiles given by (\ref{eq:TF_phi}), and the dashed black
lines are the Thomas-Fermi profiles obtained from the Gross-Pitaevskii
equation.  The parameter $x_0$ is the length scale associated with the
trap, given by $x_0 = \sqrt{\hbar/2m\omega_x}$. }
\label{fig:TFprofiles}
\end{center}
\end{figure}%
\fig{\ref{fig:TFprofiles}} shows the Thomas-Fermi wavefunctions for
a spherically symmetric condensate of $10^5$ atoms of $^{85}$Rb.  The
Thomas-Fermi solution for the atom wavefunction is similar to that
for the
structureless model for the same atom number and scattering length
(also shown in the figure), with the important difference that the
denominator has a spatial dependence.  As long as the scattering
length is moderate (left panel of figure \ref{fig:TFprofiles}), the
detuning $\varepsilon$ will be large and the molecular trapping
potential will be negligible in comparison, making the Thomas-Fermi
solution indistinguishable from the standard Thomas-Fermi solution.
However, for very large scattering lengths (right panel of figure
\ref{fig:TFprofiles}), the detuning $\varepsilon$ will be small enough
for the molecular trapping potential to be significant, making the
atom-molecule Thomas-Fermi profile different from that of the
structureless atom model.

Also, note that in (\ref{eq:TF_psi}), since the potential $V_{m}(\vec
x)$ is normally
negligible compared to $2\mu_{a}-\epsilon$, the Thomas-Fermi
solution for the atomic field is essentially of the same form as that
for the GPE at the same chemical potential.  
However, for the same
total number of atoms --- counting each molecule as two atoms --- the
chemical potential for the GPE is slightly different from that found 
in this model.  As well as this, the molecule field corresponds to 
the number of \emph{elementary} molecules, and each \emph{physical 
molecule} is a superposition of an elementary molecule and an atom 
pair, as discussed in \paperi.  Closer to the Feshbach resonance, the
proportion of atom pairs 
can become  more than 50\%, although in the systems we study here,
these effects will be so small that they can be neglected.

\begin{figure}[t]
\begin{center}
\begin{psfrags}%
\psfrag{s01}[b][b]{\color[rgb]{0,0,0}\setlength{\tabcolsep}{0pt}\begin{tabular}{c}Total atom number\end{tabular}}%
\psfrag{s02}[t][t]{\color[rgb]{0,0,0}\setlength{\tabcolsep}{0pt}\begin{tabular}{c}Radius $r$ $[x_0]$\end{tabular}}%
\psfrag{s03}[b][b]{\color[rgb]{0,0,0}\setlength{\tabcolsep}{0pt}\begin{tabular}{c}Scattering length 856$a_0$\end{tabular}}%
\psfrag{s06}[][]{\color[rgb]{0,0,0}\setlength{\tabcolsep}{0pt}\begin{tabular}{c} \end{tabular}}%
\psfrag{s07}[][]{\color[rgb]{0,0,0}\setlength{\tabcolsep}{0pt}\begin{tabular}{c} \end{tabular}}%
\psfrag{s08}[b][b]{\color[rgb]{0,0,0}\setlength{\tabcolsep}{0pt}\begin{tabular}{c}Total atom number\end{tabular}}%
\psfrag{s09}[t][t]{\color[rgb]{0,0,0}\setlength{\tabcolsep}{0pt}\begin{tabular}{c}Radius $r$ $[x_0]$\end{tabular}}%
\psfrag{s10}[b][b]{\color[rgb]{0,0,0}\setlength{\tabcolsep}{0pt}\begin{tabular}{c}Scattering length 2133$a_0$\end{tabular}}%
\psfrag{s13}[][]{\color[rgb]{0,0,0}\setlength{\tabcolsep}{0pt}\begin{tabular}{c} \end{tabular}}%
\psfrag{s14}[][]{\color[rgb]{0,0,0}\setlength{\tabcolsep}{0pt}\begin{tabular}{c} \end{tabular}}%
\psfrag{x01}[t][t]{0}%
\psfrag{x02}[t][t]{0.1}%
\psfrag{x03}[t][t]{0.2}%
\psfrag{x04}[t][t]{0.3}%
\psfrag{x05}[t][t]{0.4}%
\psfrag{x06}[t][t]{0.5}%
\psfrag{x07}[t][t]{0.6}%
\psfrag{x08}[t][t]{0.7}%
\psfrag{x09}[t][t]{0.8}%
\psfrag{x10}[t][t]{0.9}%
\psfrag{x11}[t][t]{1}%
\psfrag{x12}[t][t]{0}%
\psfrag{x13}[t][t]{5}%
\psfrag{x14}[t][t]{10}%
\psfrag{x15}[t][t]{15}%
\psfrag{x16}[t][t]{0}%
\psfrag{x17}[t][t]{5}%
\psfrag{x18}[t][t]{10}%
\psfrag{x19}[t][t]{15}%
\psfrag{v01}[r][r]{0}%
\psfrag{v02}[r][r]{0.2}%
\psfrag{v03}[r][r]{0.4}%
\psfrag{v04}[r][r]{0.6}%
\psfrag{v05}[r][r]{0.8}%
\psfrag{v06}[r][r]{1}%
\psfrag{v07}[r][r]{0}%
\psfrag{v08}[r][r]{100}%
\psfrag{v09}[r][r]{200}%
\psfrag{v10}[r][r]{300}%
\psfrag{v11}[r][r]{400}%
\psfrag{v12}[r][r]{500}%
\psfrag{v13}[r][r]{0}%
\psfrag{v14}[r][r]{100}%
\psfrag{v15}[r][r]{200}%
\psfrag{v16}[r][r]{300}%
\psfrag{v17}[r][r]{400}%
\psfrag{v18}[r][r]{500}%
\resizebox{12cm}{!}{\includegraphics{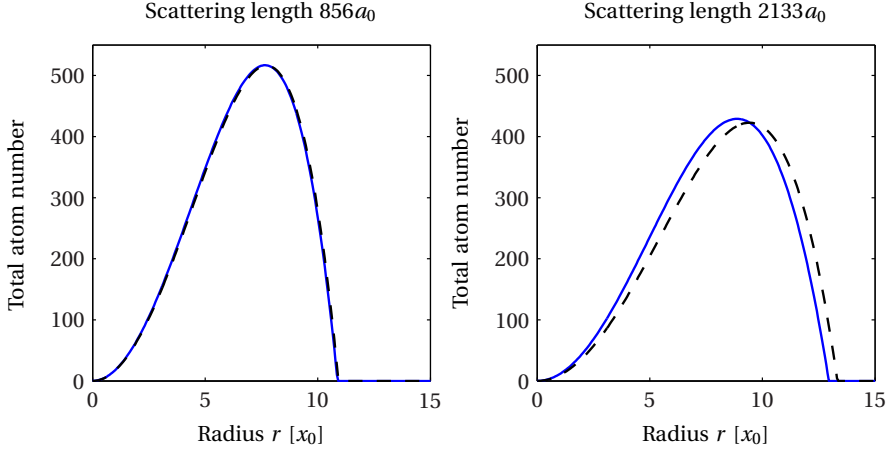}}%
\end{psfrags}%
\caption{Density for a
spherically symmetric condensate of $^{85}$Rb for two of the different
scattering lengths reported in Ref.\,\cite{claussen2003}: $a_s = 856a_0$ (left panel) and $a_s = 2133a_0$
(right panel). The system is the same as that in figure
\ref{fig:TFprofiles}. For this system, the
portion of atoms in the form of molecules is approximately $0.6\%$ for
$a_s = 856a_0$ and $4.5\%$ for $a_s = 2133a_0$. (Note: At first
glance, it appears that the solution $\psi_{GPE}$ has fewer atoms,
than given by the atom-molecule model, but, as can be seen in the
right hand figure, the discrepancy in the centre is compensated by at
the wings, where the extra factor of $4\pi r^{2}$ becomes more
significant.)}
\label{fig:TFdensities}
\end{center}
\end{figure}
The atom density of the condensate in the Thomas-Fermi limit is given
by
\begin{equation}\label{eq:TFdensity}
n\ofx = |\psi_s\ofx|^2 + 2|\phi_s\ofx|^2.
\end{equation}
\fig{\ref{fig:TFdensities}} shows the density profile for a
spherically symmetric condensate of $10^5$ atoms of $^{85}$Rb in the
Thomas-Fermi limit. For comparison we have also plotted the
Thomas-Fermi density profile obtained from a structureless atom model
for the same scattering length and atom number. 

\section{Quasiparticle excitations} 
\label{sec:Bogoliubov}%
In the Bogoliubov approximation, the field operator can be separated
into a mean-field ground state and an operator part describing the
excitations. In our coupled system, we expand the atom and molecule
wavefunctions in terms of quasiparticle bases as
\begin{eqnarray} \label{eq:BogExpansion}
\left(\begin{array}{l}\psi(\x,t) \\[3pt] \phi(\x,t) \end{array}\right)
=\left(\begin{array}{l}\psi_0\ofx \\[3pt] \phi_0\ofx
\end{array}\right) 
+\sum_{p}\left\{\left(\begin{array}{l}u_p\ofx \\[3pt]
r_p\ofx\end{array}\right)\hat{b}_p e^{-\i\omega_pt} 
+ \left(\begin{array}{l}v_p^{*}\ofx \\[3pt]
s_p^{*}\ofx\end{array}\right)\hat{b}_p^{\dagger} e^{\i\omega_pt}
\right\},
\end{eqnarray}
where the first terms are the condensate atom and molecule ground
state wavefunctions, and $\hat{b}_p$ and $\hat{b}^\dagger_p$ are the
quasiparticle destruction and creation operators respectively.  Unlike
the standard Bogoliubov expansion, where the only mixture is between
the atom creation and destruction terms, in this case it is also
necessary to include the molecule creation and destruction terms.
The quasiparticle operators in this expansion thus have both atom and
molecule components.

\subsection{Two-component Bogoliubov-de Gennes equations}
Making these substitutions to the equations of motion
(\ref{eq:EqsMotPsi}) and (\ref{eq:EqsMotPhi}) gives, after collecting
terms with the same phase, the two-component Bogoliubov-de Gennes
equations
\begin{eqnarray}
\label{eq:BdG1}
\mathcal{L}\vec U_p + \mathcal{M}\vec V_p &= & \hbar\omega_p\vec U_p,
\\ \label{eq:BdG2}
\mathcal{L}\vec V_p + \mathcal{M}^\dagger\vec U_p &=&
-\hbar\omega_p\vec V_p, 
\end{eqnarray}
where $\vec U_p = (u_p\ofx,r_p\ofx)^T$ and  $\vec V_p =
(v_p\ofx,s_p\ofx)^T$. Here $ \mathcal{L}$ is Hermitian, but
$\mathcal{M}$ need not be Hermitian.
In our case they are given by
\begin{eqnarray}\label{eq:Lgeneral}
\mathcal L &=& \begin{pmatrix}
 -{\hbar^2\nabla^2\over 2m} +V_a\ofx -\mu_a +2 U_{aa}\psi_0\ofx^2   &
g\psi_0\ofx \\[5pt]
 g\psi^*_0\ofx       &   -{\hbar^2\nabla^2\over 4m}+\varepsilon +V_m\ofx
- 2\mu_a \ \end{pmatrix},
\\[10pt] \label{eq:Mgeneral}
\mathcal M &=& \begin{pmatrix}U_{aa}\psi_0\ofx^2 + g\phi_0\ofx  & 0
\  \\[5pt]
0                   & 0  \end{pmatrix}.
\end{eqnarray}
Both $\mathcal L$ and $ \mathcal M$ are in fact Hermitian. When
$\psi_0$ and $\phi_0$ are chosen to be real, they are also symmetric.
The multi-component equations (\ref{eq:BdG1}, \ref{eq:BdG2}) are
similar to the single-component ones.  Indeed, if we let
$g\rightarrow 0$ in these equations, the upper components
obey the standard Bogoliubov-de Gennes equation for a single atom
field.

\subsection{Orthogonality and normalisation conditions}
The normalisation and orthogonality conditions of the quasiparticle
amplitudes can be derived by using the symmetry properties of
$\mathcal{L}$ and $\mathcal{M}$, for details see the appendix. We get
the conditions
\begin{eqnarray}\label{eq:Normalisation2}
\int \rmd \vec x\, \left( \vec U_{p'}^\dagger \vec U_p-  \vec
V_{p'}^\dagger \vec V_p\right) &=& \delta_{p,p'},\\ 
\label{eq:Orthogonality}
\int \rmd \vec x\, \left( \vec V_p^T \vec U_{p'} -  \vec U_p^T \vec
V_{p'}\right) &=& 0.
\end{eqnarray}
We can also use the Bose commutation relations for the components of
$\vec\psi = (\psi,\phi)^T$,
\begin{eqnarray}
\left[\vec\psi_{\alpha}\ofx,\vec\psi_{\beta}^\dagger(\x')\right]  &=&
\delta_{\alpha,\beta} \delta(\x -\x'), \\
\left[\vec\psi_{\alpha}\ofx,\vec\psi_{\beta}(\x')\right]  &=& 0,
\end{eqnarray}
from which we get the conditions
\begin{eqnarray}\label{eq:Normalisation1}
\sum_p\left( \vec U_{\alpha,p} \vec U_{\beta,p}^* - \vec
V_{\alpha,p}^* \vec V_{\beta,p} \right) &=& \delta_{\alpha,\beta} \\
\label{eq:Orthogonality2}
\sum_p\left( \vec U_{\alpha,p} \vec V_{\beta,p}^*- \vec
V_{\alpha,p}^* \vec U_{\beta,p}\right) &=& 0.
\end{eqnarray}
The four conditions (\ref{eq:Normalisation2}, \ref{eq:Orthogonality},
\ref{eq:Normalisation1}, \ref{eq:Orthogonality2}) are related; indeed
(\ref{eq:Normalisation1}) and (\ref{eq:Orthogonality2}) can be 
derived from (\ref{eq:Normalisation2}) and (\ref{eq:Orthogonality}).

\section{Uniform condensate}
If considering a uniform condensate, the Thomas-Fermi solutions
(\ref{eq:TF_psi}, \ref{eq:TF_phi}) are exact and give
\begin{eqnarray}
\label{eq:TFpsiNa}
\psi_0 &=& \sqrt{\frac{\mu_a}{U_{aa} +g^2/2(2\mu_a-\varepsilon)}}
\equiv \sqrt{n_a}, \\
\label{eq:TFphiNm}
\phi_0 &=& \frac{g}{2(2\mu_a-\varepsilon)}\psi^2  \equiv \sqrt{n_m} ,
\end{eqnarray}
where we have chosen $\psi_0$ to be real and positive.

For the uniform condensate, the quasiparticle amplitudes can be
expressed as plane waves, according to
\begin{eqnarray}
u_{\vec k,l}\ofx &\rightarrow& \frac{u_{k,l}
e^{\i\kbf\cdot\x}}{\sqrt{\mathcal{V}}} \\
v^*_{\vec k,l}\ofx &\rightarrow& \frac{v^*_{k,l}
e^{-\i\kbf\cdot\x}}{\sqrt{\mathcal{V}}}  \\
r_{\vec k,l}\ofx &\rightarrow& \frac{r_{k,l}
e^{\i\kbf\cdot\x}}{\sqrt{\mathcal{V}}}  \\
s^*_{\vec k,l}\ofx &\rightarrow& \frac{r^*_{k,l}
e^{-\i\kbf\cdot\x}}{\sqrt{\mathcal{V}}} ,
\end{eqnarray}
where $\mathcal{V}$ is the volume of the system.

In this case we can express $ \mathcal L$ and $ \mathcal M$ as
\begin{eqnarray}\label{mbdeg3}
\mathcal L &=& \begin{pmatrix}
 {\hbar^2k^2\over 2m} -\mu_a +2 U_{aa}n_a   &   g\sqrt{n_a}
\\[5pt]
   g\sqrt{n_a}          &   {\hbar^2k^2\over 4m}+\varepsilon - 2\mu_a \ 
               \end{pmatrix},
\\[10pt] \label{mbdeg4}
\mathcal M &=& \begin{pmatrix}
U_{aa}n_a + g\sqrt{n_m} & 0  \ 
\\[5pt]
0                   & 0 
\end{pmatrix}.
\end{eqnarray}
Since there will be two distinct eigenfrequencies for each value of
the momentum $\vec k$ --- one corresponding to a state which is
mainly atomic, and one corresponding to a state which is mainly
molecular, as we shall see in the following section --- the subscript
$p$ refers to the different
momentum modes $\vec k$, as well as the two eigenvalues, which we 
refer to as the  {``atomic'' $\omega_{A}$ and ``molecular''  
$\omega_{M}$.  This terminology corresponds to the behaviour of the 
eigenvectors for sufficiently small $|\vec k|$.  The situation becomes 
somewhat complicated at higher values of $|\vec k|$, as we shall show 
in the next section.}

\subsection{Eigenvalues of the Bogoliubov-de Gennes equations}
{The eigenvalues of the Bogoliubov-de Gennes equations are now given
explicitly by  a somewhat intricate procedure as follows:
\\ {\bf A : }  Define the quantities 
\begin{eqnarray}\label{Rdef}
 \mathcal{R} &\equiv & 
 \mathcal{L}_{11}^2-\mathcal{M}_{11}^2+\mathcal{L}_{22}^2,
 \\ \label{Sdef}
 \mathcal{S} &\equiv & 
 \left(\mathcal{L}_{11}^2-\mathcal{M}_{11}^2-\mathcal{L}_{22}^2\right)^2+
 4\mathcal{L}_{12}^2\left(\left(\mathcal{L}_{11}+\mathcal{L}_{22}\right)^2-\mathcal{M}_{11}^2\right).
\end{eqnarray}}
\\ {\bf B : } define the quantities $K^{l}$ and $K^{h}$ as the 
smallest and largest positive values of $|\vec k |$ for which 
$\mathcal S=0$.
\\ {\bf C : }
For the range of values of  $|\vec k|$ given by $0\le|\vec k |\le K^{l}$ , 
\begin{eqnarray}\label{LowEigenvalues1}
    \omega_{k,A}  &=& {1\over\sqrt{2}}\sqrt{\mathcal R - \sqrt{\mathcal S}},
    \\\label{LowEigenvalues2}
    \omega_{k,M}  &=& -{1\over\sqrt{2}}\sqrt{\mathcal R + 
    \sqrt{\mathcal S}}.
    \end{eqnarray}
\\ {\bf D : }
For the range of values of $|\vec k|$ given by  $K^{l}\le|\vec k |\le K^{h} $ , 
\begin{eqnarray}\label{ImagEigenvalues1}
    \omega_{k,A}  &=& {1\over\sqrt{2}}\sqrt{\mathcal R - i\sqrt{|\mathcal S|}},
    \\\label{ImagEigenvalues2}
    \omega_{k,M}  &=& -{1\over\sqrt{2}}\sqrt{\mathcal R + i
    \sqrt{|\mathcal S|}}.
    \end{eqnarray}
\\ {\bf E : }
 For the range of values of  $|\vec k|$ given by $ K^{h}\le|\vec k $ , 
\begin{eqnarray}\label{HighEigenvalues1}
    \omega_{k,A}  &=& {1\over\sqrt{2}}\sqrt{\mathcal R + \sqrt{\mathcal S}},
    \\\label{HighEigenvalues2}
    \omega_{k,M}  &=& -{1\over\sqrt{2}}\sqrt{\mathcal R - 
    \sqrt{\mathcal S}}.
    \end{eqnarray}%
\fig{\ref{fig:Eigenvalues}} shows the eigenvalues {calculated using 
this procedure} for a uniform
condensate of $^{85}$Rb with a total density of $n\equiv
n_a+2n_m=10^{20}\text{ m}^{-3}$ at a scattering length of $900a_0$.
Here we use the experimental values of the binding energy taken from
\cite{claussen2003} to determine $\varepsilon$ and $g$, using
Equations (\ref{eq:epsilon}) and (\ref{eq:g2}).  The momentum is
measured in units of the momentum $q$ of the Bragg pulse in the
experiment of Ref.  \cite{papp2008} with $q = 1.6\times 10^7$
m$^{-1}$. The following points should be noted:
\begin{figure}[t]
\begin{center}
\begin{psfrags}%
\psfrag{s01}[t][t]{\color[rgb]{0,0,0}\setlength{\tabcolsep}{0pt}\begin{tabular}{c}Momentum $k$ $[q]$\end{tabular}}%
\psfrag{s02}[b][b]{\color[rgb]{0,0,0}\setlength{\tabcolsep}{0pt}\begin{tabular}{c}Eigenvalues $\omega_{k,l}$ [kHz]\end{tabular}}%
\psfrag{s13}[lt][lt]{\color[rgb]{0,0,0}\setlength{\tabcolsep}{0pt}\begin{tabular}{l}$\omega_{GPE}$\end{tabular}}%
\psfrag{s14}[lt][lt]{\color[rgb]{0,0,0}\setlength{\tabcolsep}{0pt}\begin{tabular}{l}$\omega_{k,A}$\end{tabular}}%
\psfrag{s15}[lt][lt]{\color[rgb]{0,0,0}\setlength{\tabcolsep}{0pt}\begin{tabular}{l}$\omega_{k,M}$\end{tabular}}%
\psfrag{s16}[lt][lt]{\color[rgb]{0,0,0}\setlength{\tabcolsep}{0pt}\begin{tabular}{l}$\varepsilon$\end{tabular}}%
\psfrag{s17}[lt][lt]{\color[rgb]{0,0,0}\setlength{\tabcolsep}{0pt}\begin{tabular}{l}$-\omega_{k,A}$\end{tabular}}%
\psfrag{s18}[lt][lt]{\color[rgb]{0,0,0}\setlength{\tabcolsep}{0pt}\begin{tabular}{l}$-\omega_{k,M}$\end{tabular}}%
\psfrag{x01}[t][t]{0}%
\psfrag{x02}[t][t]{0.1}%
\psfrag{x03}[t][t]{0.2}%
\psfrag{x04}[t][t]{0.3}%
\psfrag{x05}[t][t]{0.4}%
\psfrag{x06}[t][t]{0.5}%
\psfrag{x07}[t][t]{0.6}%
\psfrag{x08}[t][t]{0.7}%
\psfrag{x09}[t][t]{0.8}%
\psfrag{x10}[t][t]{0.9}%
\psfrag{x11}[t][t]{1}%
\psfrag{x12}[t][t]{0}%
\psfrag{x13}[t][t]{0.5}%
\psfrag{x14}[t][t]{1}%
\psfrag{x15}[t][t]{1.5}%
\psfrag{x16}[t][t]{2}%
\psfrag{x17}[t][t]{2.5}%
\psfrag{v01}[r][r]{0}%
\psfrag{v02}[r][r]{0.1}%
\psfrag{v03}[r][r]{0.2}%
\psfrag{v04}[r][r]{0.3}%
\psfrag{v05}[r][r]{0.4}%
\psfrag{v06}[r][r]{0.5}%
\psfrag{v07}[r][r]{0.6}%
\psfrag{v08}[r][r]{0.7}%
\psfrag{v09}[r][r]{0.8}%
\psfrag{v10}[r][r]{0.9}%
\psfrag{v11}[r][r]{1}%
\psfrag{v12}[r][r]{-100}%
\psfrag{v13}[r][r]{-80}%
\psfrag{v14}[r][r]{-60}%
\psfrag{v15}[r][r]{-40}%
\psfrag{v16}[r][r]{-20}%
\psfrag{v17}[r][r]{0}%
\psfrag{v18}[r][r]{20}%
\psfrag{v19}[r][r]{40}%
\psfrag{v20}[r][r]{60}%
\psfrag{v21}[r][r]{80}%
\psfrag{v22}[r][r]{100}%
\resizebox{12cm}{!}{\includegraphics{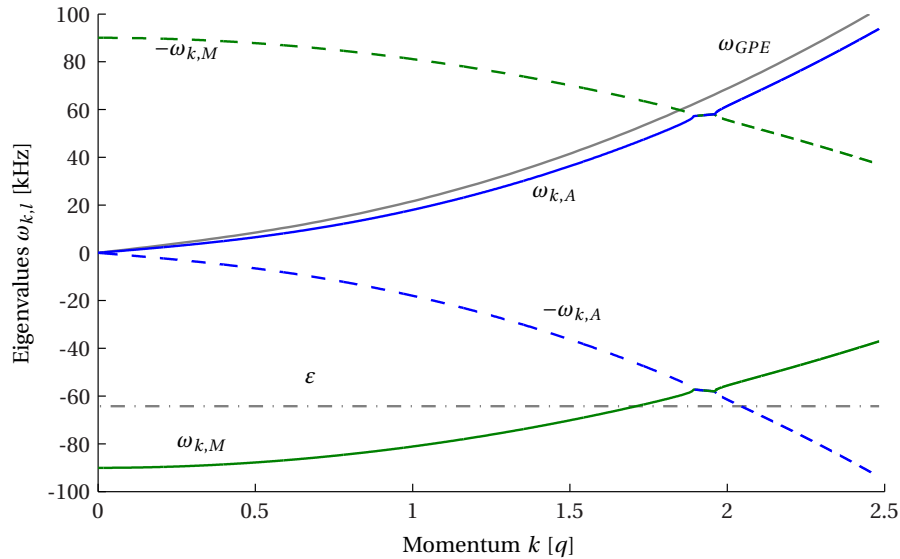}}%
\end{psfrags}%
\caption{Eigenvalues $\omega_{k,A}$ (blue solid line) and
$\omega_{k,M}$ (green solid line) as functions of the momentum $k$
for $^{85}$Rb at $900a_0$. The atom-dominated eigenvalue is similar
to that obtained from the Gross-Pitaevskii equation (grey solid
line), whereas the molecule-dominated eigenvalue is related the
molecule field detuning $\varepsilon$ (grey dash-dotted line). The
total density of the condensate is here $10^{20}$ m$^{-3}$ and $q$ is
$1.6\times 10^7$ m$^{-1}$.}
\label{fig:Eigenvalues}
\end{center}
\end{figure}
\begin{enumerate}
\item As can be seen clearly in the figure, the {atomic} eigenvalue
$\omega_{{k,A}}$ is similar to that obtained from the
Gross-Pitaevskii equation, but is slightly shifted from this, due to
the coupling between the atoms and the molecules.  This shift is not
dramatic, but is still measurable, for example by using Bragg
scattering, as in the experiment of \cite{papp2008}.

\item The atomic eigenvalue
$\omega_{{k,A}}$ belongs to the  atom-dominated
state. The energy to create an
atom-dominated quasiparticle is thus $\hbar\omega_{k,A}$. 

\item Similarly, the energy to create an molecule-dominated
quasiparticle is given by $\hbar\omega_{k,M}$.  Since
$\omega_{{k,M}}$ is related to the binding energy of the
molecules, and $\varepsilon$ is negative, $\omega_{k,M}$ is also
negative.  Thus, the energy to create a molecule-dominated
quasiparticle is \emph{negative}, and increases with $|\vec k|$, which
is expected for a bound state.

\item The higher eigenvalue $\omega_{{k,M}}$ has the same general
behaviour as $\omega_{{k,A}}$, but since it is related to the
binding energy of the molecules, it is shifted downwards.  As $k
\rightarrow 0$, and as long as the density is moderate,
$\omega_{{k,M}}$ can be approximated by
\begin{equation}
\hbar\omega_{0,H} \approx  \varepsilon+2U_{aa}n - 4\mu_a,
\end{equation}
and as $n\rightarrow 0$ it will approach the value of the molecule
field detuning $\varepsilon$.
\end{enumerate}
\subsubsection{High momentum instability}
\begin{figure}[t]
\begin{center}
\begin{psfrags}%
\psfrag{s01}[t][t]{\color[rgb]{0,0,0}\setlength{\tabcolsep}{0pt}\begin{tabular}{c}Momentum $k$ $[q]$\end{tabular}}%
\psfrag{s02}[b][b]{\color[rgb]{0,0,0}\setlength{\tabcolsep}{0pt}\begin{tabular}{c}Re$(\omega_{k,l})$ [kHz]\end{tabular}}%
\psfrag{s05}[t][t]{\color[rgb]{0,0,0}\setlength{\tabcolsep}{0pt}\begin{tabular}{c}Momentum $k$ $[q]$\end{tabular}}%
\psfrag{s06}[b][b]{\color[rgb]{0,0,0}\setlength{\tabcolsep}{0pt}\begin{tabular}{c}Im$(\omega_{k,l})$ [kHz]\end{tabular}}%
\psfrag{s13}[lt][lt]{\color[rgb]{0,0,0}\setlength{\tabcolsep}{0pt}\begin{tabular}{l}${-\omega_{k,M}}$\end{tabular}}%
\psfrag{s14}[lt][lt]{\color[rgb]{0,0,0}\setlength{\tabcolsep}{0pt}\begin{tabular}{l}${\omega_{k,A}}$\end{tabular}}%
\psfrag{s15}[lt][lt]{\color[rgb]{0,0,0}\setlength{\tabcolsep}{0pt}\begin{tabular}{l}$\omega_{GPE}$\end{tabular}}%
\psfrag{s16}[lt][lt]{\color[rgb]{0,0,0}\setlength{\tabcolsep}{0pt}\begin{tabular}{l}${-\omega_{k,M}}$\end{tabular}}%
\psfrag{s17}[lt][lt]{\color[rgb]{0,0,0}\setlength{\tabcolsep}{0pt}\begin{tabular}{l}${\omega_{k,A}}$\end{tabular}}%
\psfrag{x01}[t][t]{0}%
\psfrag{x02}[t][t]{0.1}%
\psfrag{x03}[t][t]{0.2}%
\psfrag{x04}[t][t]{0.3}%
\psfrag{x05}[t][t]{0.4}%
\psfrag{x06}[t][t]{0.5}%
\psfrag{x07}[t][t]{0.6}%
\psfrag{x08}[t][t]{0.7}%
\psfrag{x09}[t][t]{0.8}%
\psfrag{x10}[t][t]{0.9}%
\psfrag{x11}[t][t]{1}%
\psfrag{x12}[t][t]{1.8}%
\psfrag{x13}[t][t]{1.9}%
\psfrag{x14}[t][t]{2}%
\psfrag{x15}[t][t]{2.1}%
\psfrag{x16}[t][t]{1.8}%
\psfrag{x17}[t][t]{1.9}%
\psfrag{x18}[t][t]{2}%
\psfrag{x19}[t][t]{2.1}%
\psfrag{v01}[r][r]{0}%
\psfrag{v02}[r][r]{0.2}%
\psfrag{v03}[r][r]{0.4}%
\psfrag{v04}[r][r]{0.6}%
\psfrag{v05}[r][r]{0.8}%
\psfrag{v06}[r][r]{1}%
\psfrag{v07}[r][r]{-50}%
\psfrag{v08}[r][r]{0}%
\psfrag{v09}[r][r]{50}%
\psfrag{v10}[r][r]{0}%
\psfrag{v11}[r][r]{20}%
\psfrag{v12}[r][r]{40}%
\psfrag{v13}[r][r]{60}%
\psfrag{v14}[r][r]{80}%
\psfrag{v15}[r][r]{100}%
\resizebox{12cm}{!}{\includegraphics{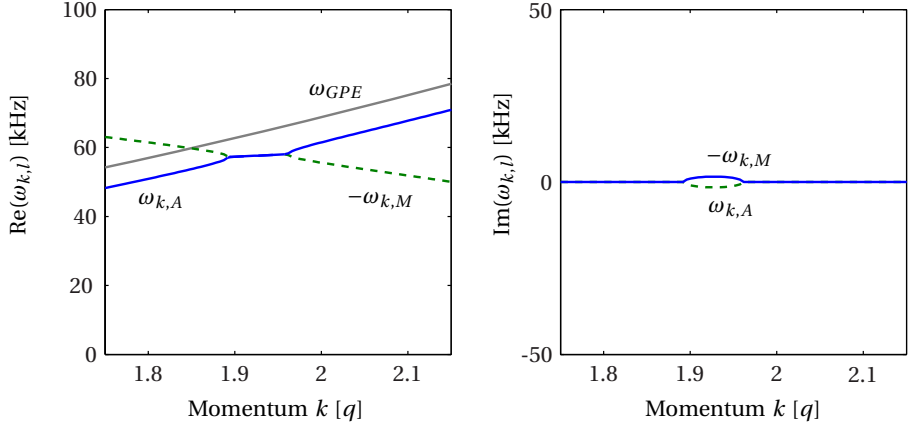}}%
\end{psfrags}%
\caption{High eigenvalue (green dashed line) and low eigenvalue
(solid blue line) for the same system as in figure
\ref{fig:Eigenvalues} in the region where the energy associated with
the destruction of an atom coincides with that of the creation of a
molecule. The eigenvalues become complex where the overlap exists,
with their real parts shown in the left panel and their respective
imaginary components are shown in the right panel. As in figure
\ref{fig:Eigenvalues}, the grey line is the eigenvalue obtained from
a system of structureless atoms.}
\label{fig:EigenvaluesDetail}
\end{center}
\end{figure}

As can be seen in \fig{\ref{fig:Eigenvalues}}, there is a crossover
which occurs when $\omega_{{k,A}}+\omega_{{k,M}}=0$, where the
energy to create simultaneously a molecular-dominated quasiparticle
and an atom-dominated quasiparticle is zero.  
{As one proceeds through the crossover, the eigenvalue formula is
determined successively by the procedures C, D and E, as given above.
The midpoint of the crossover region is at the momentum
\begin{equation}
\frac{\hbar^2k^2}{2m} = 2\mu_a -
\frac{2}{3}\left(2U_{aa}n+\varepsilon\right).
\end{equation}
\fig{\ref{fig:EigenvaluesDetail}} shows the behaviour of the high and
low eigenvalues in the crossover region, where each eigenvalue has an
imaginary part. } The maximum amplitude of the imaginary parts of the
eigenvalues at approximately $1.5$ kHz.  For this system, we would
therefore expect a relatively fast instability to occur --- on a time
scale of approximately $0.07$ ms --- with atom-molecule pairs being
created and destroyed.  

The instability arises because the energy to create a molecular
quasiparticle of momentum $\vec k$ becomes equal to the energy to
destroy an atomic quasiparticle of momentum $-\vec k$.  This is a
natural instability to expect, and provides a mechanism for the
condensate, which is metastable when the underlying interaction is
attractive, to achieve its true ground state, a condensate of
molecules.  The relatively high value of the momentum required would
make this a very much less important phenomenon in a trapped
condensate, since the wavefunctions of high energy quasiparticles are
largely located outside of the condensate.  Since the transformation
can only take place where there is a condensate, this would
significantly decrease the size of the imaginary part.

\subsection{Eigenvectors of the Bogoliubov-de Gennes equations}
\begin{figure}[t]
\begin{center}
\subfigure[$k=0.1q$]{
\psfrag{s01}[t][t]{\color[rgb]{0,0,0}\setlength{\tabcolsep}{0pt}\begin{tabular}{c}Scattering length $a_s$ $[a_0]$\end{tabular}}%
\psfrag{s02}[b][b]{\color[rgb]{0,0,0}\setlength{\tabcolsep}{0pt}\begin{tabular}{c}Eigenvector components\end{tabular}}%
\psfrag{s04}[b][b]{\color[rgb]{0,0,0}\setlength{\tabcolsep}{0pt}\begin{tabular}{c}{Atomic  eigenvalue} ${\omega_{k,A}}$\end{tabular}}%
\psfrag{s05}[t][t]{\color[rgb]{0,0,0}\setlength{\tabcolsep}{0pt}\begin{tabular}{c}Scattering length $a_s$ $[a_0]$\end{tabular}}%
\psfrag{s06}[b][b]{\color[rgb]{0,0,0}\setlength{\tabcolsep}{0pt}\begin{tabular}{c}Eigenvector components\end{tabular}}%
\psfrag{s08}[b][b]{\color[rgb]{0,0,0}\setlength{\tabcolsep}{0pt}\begin{tabular}{c}{Molecular eigenvalue} ${\omega_{k,M}}$\end{tabular}}%
\psfrag{s13}[lt][lt]{\color[rgb]{0,0,0}\setlength{\tabcolsep}{0pt}\begin{tabular}{l}$u_{{{k,A}}}$\end{tabular}}%
\psfrag{s14}[lt][lt]{\color[rgb]{0,0,0}\setlength{\tabcolsep}{0pt}\begin{tabular}{l}$r_{{{k,A}}}$\end{tabular}}%
\psfrag{s15}[lt][lt]{\color[rgb]{0,0,0}\setlength{\tabcolsep}{0pt}\begin{tabular}{l}$s_{{{k,A}}}$\end{tabular}}%
\psfrag{s16}[lt][lt]{\color[rgb]{0,0,0}\setlength{\tabcolsep}{0pt}\begin{tabular}{l}$r_{{{k,M}}}$\end{tabular}}%
\psfrag{s17}[lt][lt]{\color[rgb]{0,0,0}\setlength{\tabcolsep}{0pt}\begin{tabular}{l}$s_{{{k,M}}}$\end{tabular}}%
\psfrag{s18}[lt][lt]{\color[rgb]{0,0,0}\setlength{\tabcolsep}{0pt}\begin{tabular}{l}$u_{{{k,M}}}$\end{tabular}}%
\psfrag{s19}[lt][lt]{\color[rgb]{0,0,0}\setlength{\tabcolsep}{0pt}\begin{tabular}{l}$v_{{{k,M}}}$\end{tabular}}%
\psfrag{s20}[lt][lt]{\color[rgb]{0,0,0}\setlength{\tabcolsep}{0pt}\begin{tabular}{l}$v_{{{k,A}}}$\end{tabular}}%
\psfrag{x01}[t][t]{0}%
\psfrag{x02}[t][t]{0.1}%
\psfrag{x03}[t][t]{0.2}%
\psfrag{x04}[t][t]{0.3}%
\psfrag{x05}[t][t]{0.4}%
\psfrag{x06}[t][t]{0.5}%
\psfrag{x07}[t][t]{0.6}%
\psfrag{x08}[t][t]{0.7}%
\psfrag{x09}[t][t]{0.8}%
\psfrag{x10}[t][t]{0.9}%
\psfrag{x11}[t][t]{1}%
\psfrag{x12}[t][t]{200}%
\psfrag{x13}[t][t]{400}%
\psfrag{x14}[t][t]{600}%
\psfrag{x15}[t][t]{800}%
\psfrag{x16}[t][t]{1000}%
\psfrag{x17}[t][t]{200}%
\psfrag{x18}[t][t]{400}%
\psfrag{x19}[t][t]{600}%
\psfrag{x20}[t][t]{800}%
\psfrag{x21}[t][t]{1000}%
\psfrag{v01}[r][r]{0}%
\psfrag{v02}[r][r]{0.2}%
\psfrag{v03}[r][r]{0.4}%
\psfrag{v04}[r][r]{0.6}%
\psfrag{v05}[r][r]{0.8}%
\psfrag{v06}[r][r]{1}%
\psfrag{v07}[r][r]{-1.5}%
\psfrag{v08}[r][r]{-1}%
\psfrag{v09}[r][r]{-0.5}%
\psfrag{v10}[r][r]{0}%
\psfrag{v11}[r][r]{0.5}%
\psfrag{v12}[r][r]{1}%
\psfrag{v13}[r][r]{1.5}%
\psfrag{v14}[r][r]{-1.5}%
\psfrag{v15}[r][r]{-1}%
\psfrag{v16}[r][r]{-0.5}%
\psfrag{v17}[r][r]{0}%
\psfrag{v18}[r][r]{0.5}%
\psfrag{v19}[r][r]{1}%
\psfrag{v20}[r][r]{1.5}%
\resizebox{12cm}{!}{\includegraphics{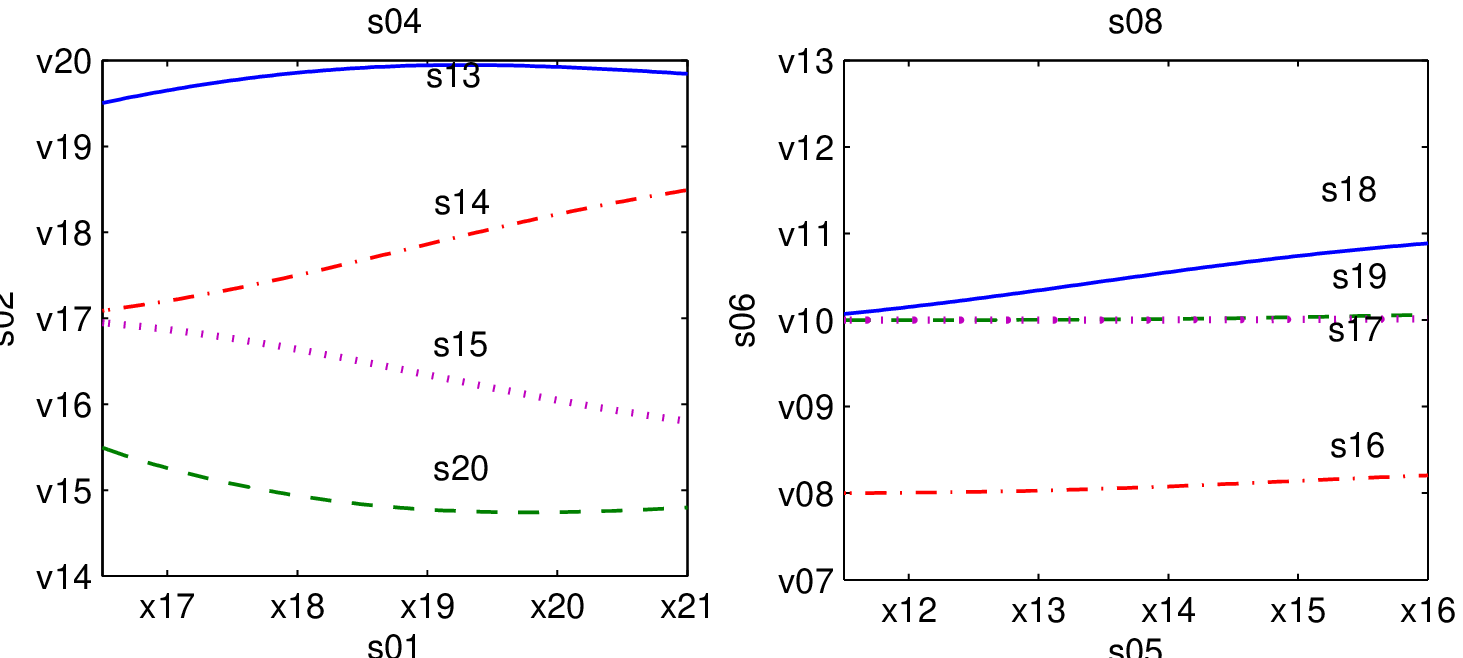}}%
\label{fig:EigenvectorsQ01}}\vspace{0.4in}
\subfigure[$k=q$]{
\psfrag{s01}[t][t]{\color[rgb]{0,0,0}\setlength{\tabcolsep}{0pt}\begin{tabular}{c}Scattering length $a_s$ $[a_0]$\end{tabular}}%
\psfrag{s02}[b][b]{\color[rgb]{0,0,0}\setlength{\tabcolsep}{0pt}\begin{tabular}{c}Eigenvector components\end{tabular}}%
\psfrag{s04}[b][b]{\color[rgb]{0,0,0}\setlength{\tabcolsep}{0pt}\begin{tabular}{c}{Atomic  eigenvalue} ${\omega_{k,A}}$\end{tabular}}%
\psfrag{s05}[t][t]{\color[rgb]{0,0,0}\setlength{\tabcolsep}{0pt}\begin{tabular}{c}Scattering length $a_s$ $[a_0]$\end{tabular}}%
\psfrag{s06}[b][b]{\color[rgb]{0,0,0}\setlength{\tabcolsep}{0pt}\begin{tabular}{c}Eigenvector components\end{tabular}}%
\psfrag{s08}[b][b]{\color[rgb]{0,0,0}\setlength{\tabcolsep}{0pt}\begin{tabular}{c}{Molecular eigenvalue} ${\omega_{k,M}}$\end{tabular}}%
\psfrag{s13}[lt][lt]{\color[rgb]{0,0,0}\setlength{\tabcolsep}{0pt}\begin{tabular}{l}$r_{{{k,A}}}$\end{tabular}}%
\psfrag{s14}[lt][lt]{\color[rgb]{0,0,0}\setlength{\tabcolsep}{0pt}\begin{tabular}{l}$u_{{{k,A}}}$\end{tabular}}%
\psfrag{s15}[lt][lt]{\color[rgb]{0,0,0}\setlength{\tabcolsep}{0pt}\begin{tabular}{l}$s_{{{k,M}}}$\end{tabular}}%
\psfrag{s16}[lt][lt]{\color[rgb]{0,0,0}\setlength{\tabcolsep}{0pt}\begin{tabular}{l}$u_{{{k,M}}}$\end{tabular}}%
\psfrag{s17}[lt][lt]{\color[rgb]{0,0,0}\setlength{\tabcolsep}{0pt}\begin{tabular}{l}$r_{{{k,M}}}$\end{tabular}}%
\psfrag{s18}[lt][lt]{\color[rgb]{0,0,0}\setlength{\tabcolsep}{0pt}\begin{tabular}{l}$v_{{{k,M}}}$\end{tabular}}%
\psfrag{s19}[lt][lt]{\color[rgb]{0,0,0}\setlength{\tabcolsep}{0pt}\begin{tabular}{l}$v_{{{k,A}}}$\end{tabular}}%
\psfrag{s20}[lt][lt]{\color[rgb]{0,0,0}\setlength{\tabcolsep}{0pt}\begin{tabular}{l}$s_{{{k,A}}}$\end{tabular}}%
\psfrag{x01}[t][t]{0}%
\psfrag{x02}[t][t]{0.1}%
\psfrag{x03}[t][t]{0.2}%
\psfrag{x04}[t][t]{0.3}%
\psfrag{x05}[t][t]{0.4}%
\psfrag{x06}[t][t]{0.5}%
\psfrag{x07}[t][t]{0.6}%
\psfrag{x08}[t][t]{0.7}%
\psfrag{x09}[t][t]{0.8}%
\psfrag{x10}[t][t]{0.9}%
\psfrag{x11}[t][t]{1}%
\psfrag{x12}[t][t]{200}%
\psfrag{x13}[t][t]{400}%
\psfrag{x14}[t][t]{600}%
\psfrag{x15}[t][t]{800}%
\psfrag{x16}[t][t]{1000}%
\psfrag{x17}[t][t]{200}%
\psfrag{x18}[t][t]{400}%
\psfrag{x19}[t][t]{600}%
\psfrag{x20}[t][t]{800}%
\psfrag{x21}[t][t]{1000}%
\psfrag{v01}[r][r]{0}%
\psfrag{v02}[r][r]{0.2}%
\psfrag{v03}[r][r]{0.4}%
\psfrag{v04}[r][r]{0.6}%
\psfrag{v05}[r][r]{0.8}%
\psfrag{v06}[r][r]{1}%
\psfrag{v07}[r][r]{-1.5}%
\psfrag{v08}[r][r]{-1}%
\psfrag{v09}[r][r]{-0.5}%
\psfrag{v10}[r][r]{0}%
\psfrag{v11}[r][r]{0.5}%
\psfrag{v12}[r][r]{1}%
\psfrag{v13}[r][r]{1.5}%
\psfrag{v14}[r][r]{-1.5}%
\psfrag{v15}[r][r]{-1}%
\psfrag{v16}[r][r]{-0.5}%
\psfrag{v17}[r][r]{0}%
\psfrag{v18}[r][r]{0.5}%
\psfrag{v19}[r][r]{1}%
\psfrag{v20}[r][r]{1.5}%
\resizebox{12cm}{!}{\includegraphics{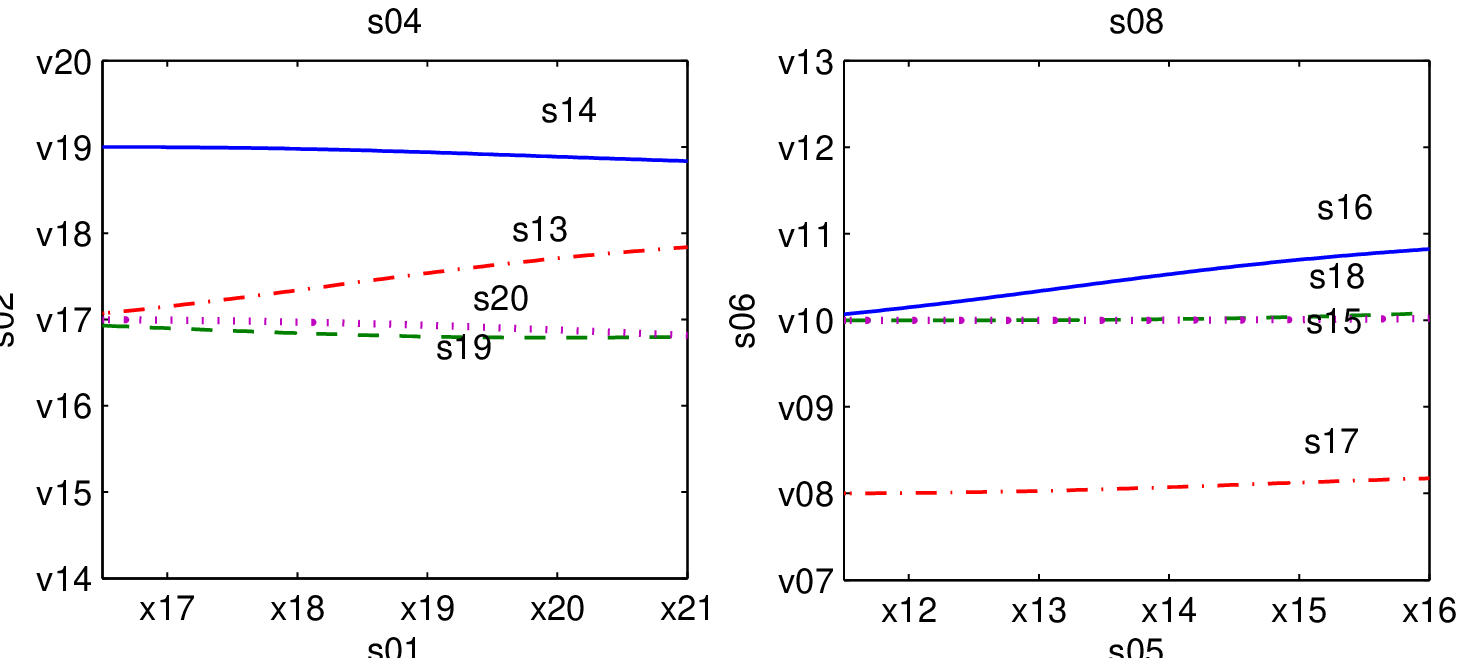}}%
\label{fig:EigenvectorsQ1}}
\caption{Eigenvector components $u_{{k,A}}$ (blue solid line),
$v_{{k,A}}$ (green dashed line), $r_{{k,A}}$ (red dash-dotted line) and
$s_{{k,A}}$ (cyan dotted line) for the low eigenvalue $\omega_{{k,A}}$
(left panel) and the high eigenvalue $\omega_{{k,M}}$ (right panel).
The momentum is
measured in units of the momentum $q$ of the Bragg pulse in the
experiment of Ref. \cite{papp2008} with $q = 1.6\times 10^7$
m$^{-1}$.}
\label{fig:Eigenvectors}
\end{center}
\end{figure}

Using the normalisation condition (\ref{eq:Normalisation1}) we can
find a unique set of eigenvectors of the Bogoliubov-de Gennes
equations corresponding to the eigenvalues 
(\ref{LowEigenvalues1}--\ref{HighEigenvalues2}):
\begin{eqnarray} \label{eq:uL}
u_{{k,A}} &=&
\sqrt{\frac{1-A_{{k,M}}^2(1-C_{{k,M}}^2)-B_{{k,M}}^2}{(B_{{k,A}}^2-A_{{k,A}}^2C_{{k,A}}^2)(1-A_{{k,M}}^2)-(B_{{k,M}}^2-A_{{k,M}}^2C_{{k,M}}^2)(1-A_{{k,A}}^2)}},
\\ \label{eq:uH}
u_{{k,M}} &=& \sqrt{\frac{1-(1-A_{{k,A}}^2)u_{{k,A}}^2}{1-A_{{k,M}}^2}}, 
\end{eqnarray}
and, for $l=L,H$,
\begin{eqnarray} \label{eq:vLH}
v_{{k,A}} &=& A_{{k,A}}u_{{k,A}}, \\ \label{eq:rLH}
r_{{k,A}} &=& B_{{k,A}}u_{{k,A}},  \\ \label{eq:sLH}
s_{{k,A}} &=& C_{{k,A}}v_{{k,A}},
\end{eqnarray}
where
\begin{eqnarray}
A_{{k,A}} &=& \frac{1}{\mathcal{M}_{11}}\left(-\mathcal{L}_{11}
+\hbar\omega_{{k,A}} -
\frac{\mathcal{L}_{12}^2}{\hbar\omega_{{k,A}}-\mathcal{L}_{22}}\right),
\\
B_{{k,A}} &=&
\frac{\mathcal{L}_{12}}{\hbar\omega_{{k,A}}-\mathcal{L}_{22}}, \\
C_{{k,A}} &=&
-\frac{\mathcal{L}_{12}}{\hbar\omega_{{k,A}}+\mathcal{L}_{22}}.
\end{eqnarray}
The components of the eigenvectors for the same system as that in
\fig{\ref{fig:Eigenvalues}} are plotted in \fig{\ref{fig:Eigenvectors}} as a function of the scattering length, for low
momentum (\fig{\ref{fig:EigenvectorsQ01}}) and moderate momentum
(\fig{\ref{fig:EigenvectorsQ1}}).  At low scattering lengths, the
eigenvector corresponding to the low eigenvalue $\omega_{{k,A}}$ is
dominated by the atom components $u_{{k,A}}$ and $v_{{k,A}}$, with the
molecule components $r_{{k,A}}$ and $s_{{k,A}}$ becoming more significant
the larger the scattering length. For larger scattering lengths, the
amplitude of the molecule
destruction operator is clearly significant compared to the atom one,
and there is a large portion of molecules in the atom state.  

For the higher eigenvalue $\omega_{{k,M}}$, the situation is reversed
with the
dominating eigenvector component being the one associated with the
molecule destruction operator.  Here we also see increased mixing
between the two states as the scattering length is increased.

\section{Application to Bragg scattering}
The quasiparticle description of the excitations in our atom-molecule
system can be used to describe the excitation spectrum from Bragg
scattering. Following the procedure of \cite{blakie2002} and
\cite{tozzo2003}, we derive the momentum transferred onto the
condensate from an Bragg pulse, and compare the results we obtain
here to those in \cite{blakie2002}.

The equations of motion for the coupled atom and molecule fields
subject to a Bragg pulse are given by
\begin{eqnarray}
\label{eq:DpsiDt}
i\hbar\frac{\partial{\psi}(\x)}{\partial t} &=&
\left(-\frac{\hbar^{2}\nabla^{2}}{2m}+V_{a}(\x)\right){\psi}(\x)+
U_{aa}|{\psi}(\x)|^{2}{\psi}(\x)+ \hbar V_{\text{opt}} \psi +
g{\psi}(\x)^{*}{\phi}(\x), \\
\label{eq:DphiDt}
i\hbar\frac{\partial{\phi}(\x)}{\partial t} &=&
\left(-\frac{\hbar^{2}\nabla^{2}}{4m}+\varepsilon+V_{m}(\x)\right){\phi}(\x)+2\hbar
V_{\text{opt}} \cos(\q\cdot\x-\omega t)\phi +\frac{g}{2}{\psi}^2(\x),
\end{eqnarray}
where
\begin{equation}
V_{\text{opt}} = V_0(t) \cos(\q\cdot\x-\omega t),
\end{equation} 
where $\q$ and $\omega$ are the wavevector and the frequency of the Bragg pulse respectively \cite{blakie2000,blakie2002}, and $V_0$ is the amplitude of the optical potential, given in terms of the Rabi frequency $\Omega$ and the excited state detuning $\Delta$,
\begin{equation}
V_0 = \frac{\hbar\Omega^2}{2\Delta}.
\end{equation}
The optical potential for the molecule is chosen to be twice that of the atom on the assumption that the atoms in the molecule are very weakly bound, and for these purposes behave almost independently.

\subsection{Quasiparticle evolution}
Using the expansion (\ref{eq:BogExpansion}), and the Bogoliubov-de
Gennes equations, we can write the equations of motion
\begin{equation}
\i\sum_p\left(\vec U_p \dot b_p e^{-\i\omega_p t} +\vec V_p^* \dot
b^*_p e^{\i\omega_p t}  \right) = \mathcal{V}_{\text{opt}}\vec\psi,
\end{equation}
where 
\begin{equation}
\mathcal{V}_{\text{opt}} = V_{\text{opt}}  \begin{pmatrix}\, 1  &   0
\\[5pt]    0   &   2 \, \end{pmatrix}.
\end{equation}
Similarly to the method in \cite{morgan1998}, we can project out
$b_p$ from the above equation by writing
\begin{align}
&\int \rmd\x\,\left\{ \left( \vec U_{p'}^\dagger
\mathcal{V}_{\text{opt}}\vec\psi + \vec
V_{p'}^\dagger\mathcal{V}_{\text{opt}}\vec\psi^*   \right)   \right\}
= \\
&\quad = \int \rmd\x\,\sum_p\left\{\i\vec U_{p'}^\dagger\left(\vec U_p
\dot b_p e^{-\i\omega_p t} +\vec V_p^* \dot b^*_p e^{\i\omega_p t}
\right)  - \i\vec V_{p'}^\dagger\left(\vec U_p^* \dot b_p^*
e^{\i\omega_p t} +\vec V_p \dot b_p e^{-\i\omega_p t}  \right)
\right\},  \\
&\quad = \i \sum_p\left\{ \int \rmd\x\, \left( \vec U_{p'}^\dagger \vec
U_p-  \vec V_{p'}^\dagger \vec V_p\right) \dot b_p e^{-\i\omega_p t}
+ \int \rmd\x\, \left( \vec U_{p'}^\dagger \vec V_{p}^* -  \vec
V_{p'}^\dagger \vec U_{p}^*\right) b^*_p e^{\i\omega_p t} \right\} ,
\\
&\quad = \i\dot{b}_{p'}(t)e^{-\i\omega_{p'}t},
\end{align}
where on the last line we have used the normalisation and
orthogonality conditions (\ref{eq:Normalisation2},
\ref{eq:Orthogonality}). We therefore have
\begin{equation}
\dot{b}_p(t) = -\i e^{i\omega_p t} \int \rmd\x\,\left\{\left( \vec
U_{p}^\dagger \mathcal{V}_{\text{opt}}\vec\psi + \vec
V_{p}^\dagger\mathcal{V}_{\text{opt}}\vec\psi^*  \right)   \right\},
\end{equation}
Setting $\vec\psi=(\psi_0,\phi_0)^T$, we get
\begin{equation}
b_p(t) = -\i \int_0^{t}e^{i\omega_p t'} dt' \int
\rmd\x\,\left\{V_{\text{opt}}  \left(
\psi_0(u_p^*+v_p^*)+2\phi_0(r_p^*+s_p^*) \right)   \right\}.
\end{equation}
Assuming that the Bragg pulse is square, with amplitude $V_0$ and
duration $T$, we get
\begin{eqnarray}
b_p(T) &=& -\frac{\i V_0}{2} \int \rmd\x\, \left(
\psi_0(u_p^*+v_p^*)+2\phi_0(r_p^*+s_p^*) \right) \nonumber\\
&& \times \int_0^{T}dt' e^{i\omega_pt'} \left(e^{\i(\q\cdot\x-\omega
t') }+e^{-\i(\q\cdot\x-\omega t' )}\right), \\
&=& -\i V_0 e^{\i\omega_pT/2}\int \rmd\x\,  \left(
\psi_0(u_p^*+v_p^*)+2\phi_0(r_p^*+s_p^*) \right) \nonumber\\
&& \times \left[ e^{-\i\omega T/2}
e^{\i\q\cdot\x}\frac{\sin\left({(\omega_p-\omega)T/2}\right)}{\omega_p-\omega}+e^{\i\omega
T/2}
e^{-\i\q\cdot\x}\frac{\sin\left({(\omega_p+\omega)T/2}\right)}{\omega_p+\omega}
\right].
\end{eqnarray}
%

\subsection{Momentum transfer from uniform condensate}
If considering the ideal case of Bragg scattering from a uniform
condensate, the quasiparticle amplitudes are given by (\ref{eq:uL} --
\ref{eq:sLH}), and the ground state is given by (\ref{eq:TF_psi},
\ref{eq:TF_phi}). The expression for $b_p$ now simplifies to
\begin{eqnarray}
b_p(T) &=& - \frac{\i V_0}{\sqrt{\mathcal{V}}}e^{\i\omega_{{k,A}}T/2}
\int_\mathcal{V}d{\x}
\left(\sqrt{n_a}(u_p^*+v_p^*)+2\sqrt{n_m}(r_p^*+s_p^*)
\right)e^{\i{\kbf}\cdot{\x}} \nonumber \\
&&\times\left[e^{-\i\omega
T/2}\frac{\sin{((\omega_{p}-\omega)T/2)}}{\omega_{p}-\omega}
e^{\i{\q}\cdot{\x}}+ e^{\i\omega
T/2}\frac{\sin{((\omega_{p}+\omega)T/2)}}{\omega_{p}+\omega}
e^{-\i{\q}\cdot{\x}}\right] ,\\
\label{eq:BogB}
&=& -\i V_0 \left( \sqrt{N_a}(u_p^*+v_p^*)+2\sqrt{N_m}(r_p^*+s_p^*)
\right) \nonumber \\
&&\times\left[e^{\i(\omega_{p}-\omega) T/2}
\frac{\sin{((\omega_{p}-\omega)T/2)}}{\omega_{p}-\omega}\delta_{p,-q}
+ e^{\i(\omega_{p}+\omega) T/2}
\frac{\sin{(\omega_{p}+\omega)T/2)}}{\omega_{p}+\omega}\delta_{p,q}
\right], \nonumber \\
\end{eqnarray}
where $N_a$ and $N_m $ are the number of atoms and molecules,
respectively.

The total momentum imparted to the condensate can now be evaluated
using the Bogoliubov expansions (\ref{eq:BogExpansion}), and the
normalisation and orthogonality relations (\ref{eq:Normalisation2},
\ref{eq:Orthogonality}, \ref{eq:Normalisation1}),
\begin{eqnarray}
P(T) &=& \frac{\hbar}{2\i}\int{\rmd\x\, \left\{\psi^*\nabla\psi +
\frac{1}{2}\phi^*\nabla\phi\right\} }+\mbox{ c.c.} \nonumber \\
&=& \sum_{p} \hbar p  \left( \left| b_{p}(T) \right|^2 +  \frac{1}{2}
\left| b_{p}(T) \right|^2   \right) \nonumber \\
&=& \frac{3\hbar qV_0^2}{2}\sum_l\left[
\left(\frac{\sin{((\omega_{q,l}-\omega)T/2)}}{\omega_{q,l}-\omega}\right)
^2  -
\left(\frac{\sin{((\omega_{q,l}+\omega)T/2)}}{\omega_{q,l}+\omega}\right)
^2  \right] \nonumber \\ \label{eq:MomentumImparted}
&& \times \left| \sqrt{N_a}(u_q+v_q)+2\sqrt{N_m}(r_q+s_q) \right|^2.
\end{eqnarray}
It is clear that this expression will have its maximum values at the
points where $\omega = \omega_{q,l}$, and if the atom field is much
larger than the molecular field --- as is the case for the system
studied here --- the only significant maximum will be at the lower
eigenvalue $\omega_L$. Consequently the Bragg resonance peak will
occur at the eigenvalue of the atom state, $\omega_{k,A}$ in
figure~\ref{fig:Eigenvalues}, corresponding to the momentum of the
Bragg pulse.

\begin{figure}[t]
\begin{center}
\begin{psfrags}
\psfrag{s05}[t][t]{\color[rgb]{0,0,0}\setlength{\tabcolsep}{0pt}\begin{tabular}{c}Bragg frequency [kHz]\end{tabular}}%
\psfrag{s06}[b][b]{\color[rgb]{0,0,0}\setlength{\tabcolsep}{0pt}\begin{tabular}{c}Fraction of scattered particles\end{tabular}}%
\psfrag{s10}[][]{\color[rgb]{0,0,0}\setlength{\tabcolsep}{0pt}\begin{tabular}{c} \end{tabular}}%
\psfrag{s11}[][]{\color[rgb]{0,0,0}\setlength{\tabcolsep}{0pt}\begin{tabular}{c} \end{tabular}}%
\psfrag{s12}[l][l]{\color[rgb]{0,0,0}$a_s = 1000a_0$}%
\psfrag{s13}[l][l]{\color[rgb]{0,0,0}$a_s = 100a_0$}%
\psfrag{s14}[l][l]{\color[rgb]{0,0,0}$a_s = 700a_0$}%
\psfrag{s15}[l][l]{\color[rgb]{0,0,0}$a_s = 1000a_0$}%
\psfrag{x01}[t][t]{0}%
\psfrag{x02}[t][t]{0.1}%
\psfrag{x03}[t][t]{0.2}%
\psfrag{x04}[t][t]{0.3}%
\psfrag{x05}[t][t]{0.4}%
\psfrag{x06}[t][t]{0.5}%
\psfrag{x07}[t][t]{0.6}%
\psfrag{x08}[t][t]{0.7}%
\psfrag{x09}[t][t]{0.8}%
\psfrag{x10}[t][t]{0.9}%
\psfrag{x11}[t][t]{1}%
\psfrag{x12}[t][t]{-30}%
\psfrag{x13}[t][t]{-20}%
\psfrag{x14}[t][t]{-10}%
\psfrag{x15}[t][t]{0}%
\psfrag{x16}[t][t]{10}%
\psfrag{x17}[t][t]{20}%
\psfrag{x18}[t][t]{30}%
\psfrag{v01}[r][r]{0}%
\psfrag{v02}[r][r]{0.1}%
\psfrag{v03}[r][r]{0.2}%
\psfrag{v04}[r][r]{0.3}%
\psfrag{v05}[r][r]{0.4}%
\psfrag{v06}[r][r]{0.5}%
\psfrag{v07}[r][r]{0.6}%
\psfrag{v08}[r][r]{0.7}%
\psfrag{v09}[r][r]{0.8}%
\psfrag{v10}[r][r]{0.9}%
\psfrag{v11}[r][r]{1}%
\psfrag{v12}[r][r]{-0.1}%
\psfrag{v13}[r][r]{-0.05}%
\psfrag{v14}[r][r]{0}%
\psfrag{v15}[r][r]{0.05}%
\psfrag{v16}[r][r]{0.1}%
\psfrag{v17}[r][r]{0.15}%
\resizebox{12cm}{!}{\includegraphics{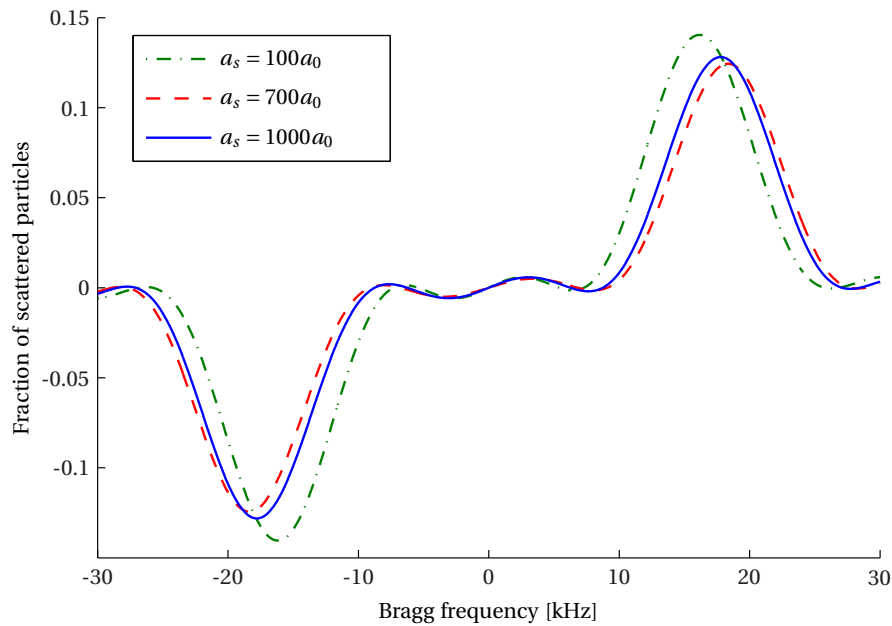}}%
\end{psfrags}%
\caption{Bragg spectrum showing the fraction of particles scattered
out of the condensate as a function of the Bragg frequency for a
uniform system $^{85}$Rb with total density of $10^{20}$ m$^{-3}$.
Here we show the spectra for three different scattering lengths:
$a_s=100a_0$ (green dashed-dotted line), $a_s=700a_0$ (red dashed
line) and $a_s=1000a_0$ (blue solid line). The duration of the Bragg
pulse is $T=0.1$ms, and the intensity of the pulse is $V_0=2\pi
\times1$ kHz.}
\label{fig:Spectra}
\end{center}
\end{figure}

\fig{\ref{fig:Spectra}} shows the fraction of particles being
scattered out of the condensate during the Bragg pulse for three
different scattering lengths. The portion of scattered particles is
related to the momentum transfer as $P/\hbar qN$, where $N \equiv N_a
+2N_m$ is the total number of particles. 
The resonance peak of the Bragg spectrum shifts to higher frequencies
as the scattering length is increased. However, this shift starts to
decrease when the scattering length is larger than approximately
$700a_0$. The free particle resonance is located at $1/2\pi \times
\hbar q^2/2m \approx 15.4$ kHz.

\begin{figure}[t]
\begin{center}
\begin{psfrags}%
\psfrag{s05}[t][t]{\color[rgb]{0,0,0}\setlength{\tabcolsep}{0pt}\begin{tabular}{c}Scattering length $a_s$ $[a_0]$\end{tabular}}%
\psfrag{s06}[b][b]{\color[rgb]{0,0,0}\setlength{\tabcolsep}{0pt}\begin{tabular}{c}Lineshift [kHz]\end{tabular}}%
\psfrag{x01}[t][t]{0}%
\psfrag{x02}[t][t]{0.1}%
\psfrag{x03}[t][t]{0.2}%
\psfrag{x04}[t][t]{0.3}%
\psfrag{x05}[t][t]{0.4}%
\psfrag{x06}[t][t]{0.5}%
\psfrag{x07}[t][t]{0.6}%
\psfrag{x08}[t][t]{0.7}%
\psfrag{x09}[t][t]{0.8}%
\psfrag{x10}[t][t]{0.9}%
\psfrag{x11}[t][t]{1}%
\psfrag{x12}[t][t]{0}%
\psfrag{x13}[t][t]{200}%
\psfrag{x14}[t][t]{400}%
\psfrag{x15}[t][t]{600}%
\psfrag{x16}[t][t]{800}%
\psfrag{x17}[t][t]{1000}%
\psfrag{v01}[r][r]{0}%
\psfrag{v02}[r][r]{0.1}%
\psfrag{v03}[r][r]{0.2}%
\psfrag{v04}[r][r]{0.3}%
\psfrag{v05}[r][r]{0.4}%
\psfrag{v06}[r][r]{0.5}%
\psfrag{v07}[r][r]{0.6}%
\psfrag{v08}[r][r]{0.7}%
\psfrag{v09}[r][r]{0.8}%
\psfrag{v10}[r][r]{0.9}%
\psfrag{v11}[r][r]{1}%
\psfrag{v12}[r][r]{0}%
\psfrag{v13}[r][r]{1}%
\psfrag{v14}[r][r]{2}%
\psfrag{v15}[r][r]{3}%
\psfrag{v16}[r][r]{4}%
\psfrag{v17}[r][r]{5}%
\psfrag{v18}[r][r]{6}%
\psfrag{v19}[r][r]{7}%
\resizebox{12cm}{!}{\includegraphics{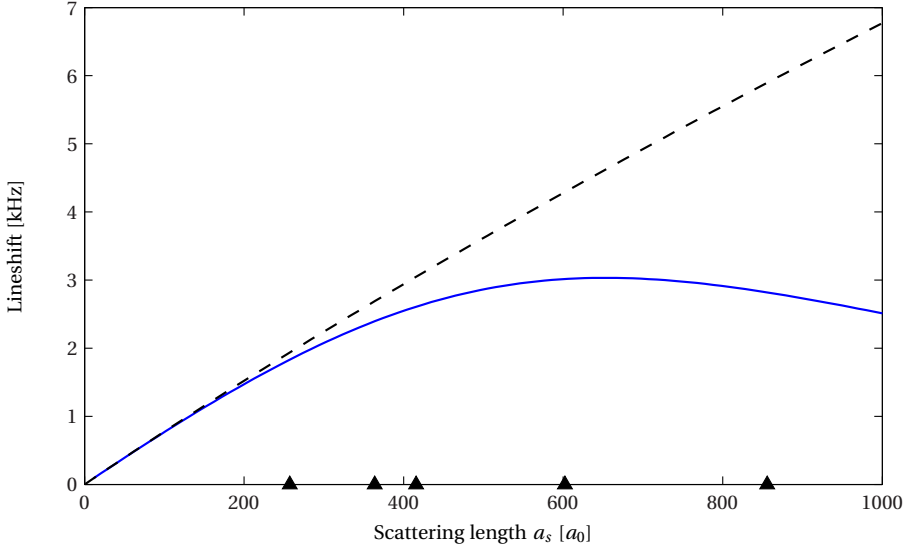}}%
\end{psfrags}%

\caption{The shift of the peak of the Bragg spectra from that of the
non-interacting gas, plotted against the scattering length $a_s$.  The
solid blue line is based on the peak position calculated by
(\ref{eq:MomentumImparted}), and the black dashed line is the
equivalent calculation based on the model of structureless atoms, as
in \cite{blakie2002}.  The black triangles mark the values of the
scattering length for which Ref.  \cite{claussen2003} gives
experimental data of the binding energy.  }
\label{fig:Lineshift}
\end{center}
\end{figure}

\fig{\ref{fig:Lineshift}} shows how the shift of the Bragg resonance
from the free particle resonance changes with the increase in the
scattering length.  For comparison we have also plotted the result
based on the model of a condensate of structureless atoms with
increased scattering length \cite{blakie2002}. The two models agree
well for low scattering lengths, but differ dramatically for larger
scattering lengths. In the figure we have also marked the scattering
lengths for which there is available data of the binding energy, as
in \cite{claussen2003}. The values outside of these are calculated by
the method described in Section\,\ref{sec:formalism}.

\section{Conclusion}
The Bragg peak shift obtained from our model of a coupled
atom-molecule condensate is significantly different from that
obtained by a model of structureless atoms. Even though it describes
the ideal case of Bragg scattering from a uniform condensate and is
therefore not directly comparable to that obtained in the experiment
of Ref. \cite{papp2008}, it is still qualitatively similar to this.
Most importantly, unlike other approaches to this problem, for
example that by Kinnunen \etal \cite{kinnunen2009}, the lineshift in
figure~\ref{fig:Lineshift} shows a clear \emph{downward} behaviour
for large scattering lengths, exactly as was reported experimentally in \cite{papp2008}.

In \paperiii\, we will implement the formalism described here and in
\paperi\ in full numerical calculations of Bragg scattering from an
inhomogeneous BEC. The results from these simulations will be
directly comparable to the results from the experiment of
\cite{papp2008}.

\appendix
\section{Orthogonality and normalization conditions}
The orthogonality conditions of the quasiparticle amplitudes can be
derived by writing
\begin{eqnarray}
\hbar\omega_p \int \rmd \vec x\, \left( \vec U_{p'}^\dagger \vec U_p
-  \vec V_{p'}^\dagger \vec V_p\right)
&=& \int \rmd \vec x\,\vec U_{p'}^\dagger\Big(\mathcal L \vec U_p +
\mathcal M \vec V_p\Big)
+ \int \rmd \vec x\,\vec V_{p'}^\dagger\left( \mathcal L \vec V_p +
\mathcal M^\dagger \vec U_p\right)\nonumber \\
\\ \label{mbdeg6}
\hbar\omega_{p'} \int \rmd \vec x\, \left( \vec U_{p'}^\dagger \vec
U_p -  \vec V_{p'}^\dagger \vec V_p\right)
&=&\int \rmd \vec x\, \left(\vec U^\dagger_{p'}\mathcal L  +  \vec
V^\dagger_{p'}\mathcal M^\dagger\right)\vec U_p
+ \int \rmd \vec x\,\left(\vec V_{p'}^\dagger\mathcal L^\dagger + \vec
U^\dagger_{p'}\mathcal M \right)\vec V_p .
\nonumber\\
\end{eqnarray}
The Hermitian nature of $ \mathcal L$ means that the two right hand
sides are 
equal so that, for an appropriate normalization we can write
\begin{eqnarray}
\int \rmd \vec x\, \left( \vec U_{p'}^\dagger \vec U_p-  \vec
V_{p'}^\dagger \vec V_p\right) &=& \delta_{p,p'}.
\end{eqnarray}
Since both $\mathcal{L}$ and $\mathcal{M}$ are symmetric we can then
write
\begin{eqnarray}\label{mbdeg8}
\hbar\omega_{p'} \int \rmd \vec x\, \left( \vec V_p^T \vec U_{p'}  -
\vec U_p^T \vec V_{p'}\right)
&=& \int \rmd \vec x\,  \left\{\vec V_p^T\left(\mathcal{L}\vec U_{p'} +
\mathcal{M}\vec V_{p'}\right)
+ \vec U_p^T\left(\mathcal{L}\vec V_{p'} + \mathcal{M}^\dagger\vec
U_{p'}\right)\right\} \nonumber \\
\\ \label{mbdeg9}
\hbar\omega_p \int \rmd \vec x\, \left( \vec V_p^T \vec U_{p'}  -
\vec U_p^T \vec V_{p'}\right)
&=&-\int \rmd \vec x\, \left\{\left(\vec V_p^T   \mathcal{L} + \vec
U_p^T\mathcal{M}^*\right)\vec U_{p'}
- \left(\vec U_p^T\mathcal{L} + \vec V_p\mathcal{M}\right)\vec V_{p'}\right\},
\nonumber\\
\end{eqnarray}
so that, assuming $\omega_p\neq -\omega_{p'}$, we can say
\begin{eqnarray}
\int \rmd \vec x\, \left( \vec V_p^T \vec U_{p'} -  \vec U_p^T \vec
V_{p'}\right) =0.
\end{eqnarray}

We can also find normalisation conditions using the Bose commutation
relations for the components of $\vec\psi = (\psi,\phi)^T$,
\begin{eqnarray}
\delta_{\alpha,\beta} \delta(\x -\x') &=&
\left[\vec\psi_{\alpha}\ofx,\vec\psi_{\beta}^\dagger(\x')\right] \\
&=& \sum_{p,p'}\left\{\vec U_{\alpha,p}\ofx \vec
U_{\beta,p'}^*(\x')\left[\hat b_p\ofx,\hat b_{p'}^\dagger (\x')
\right] +\vec V_{\alpha,p}^*\ofx \vec V_{\beta,p'}(\x')\left[\hat
b_p^\dagger\ofx,\hat b_{p'}(\x')  \right]  \right. \nonumber \\
&& \left. +  \vec U_{\alpha,p}\ofx \vec V_{\beta,p'}(\x')\left[\hat
b_p\ofx,\hat b_{p'}(\x') \right] +\vec V_{\alpha,p}^*\ofx \vec
U_{\beta,p'}^*(\x')\left[\hat b_p^\dagger\ofx,\hat
b_{p'}^\dagger(\x')  \right]   \right\}
\end{eqnarray}
from which, assuming that the quasiparticle operators $\hat b_p$ obey
the usual Bose commutation relations, we get the normalisation
condition of the amplitudes,
\begin{equation}
\sum_p\left( \vec U_{\alpha,p} \vec U_{\beta,p}^* - \vec
V_{\alpha,p}^* \vec V_{\beta,p} \right) = \delta_{\alpha,\beta}
\end{equation}
Similarly, the other commutation relation,
\begin{eqnarray}
0 &=& \left[\vec\psi_{\alpha}\ofx,\vec\psi_{\beta}(\x')\right] \\
&=& \sum_{p,p'}\left\{\vec U_{\alpha,p}\ofx \vec
U_{\beta,p'}(\x')\left[\hat b_p\ofx,\hat b_{p'} (\x') \right] +\vec
V_{\alpha,p}^*\ofx \vec V_{\beta,p'}^*(\x')\left[\hat
b_p^\dagger\ofx,\hat b_{p'}^\dagger(\x')  \right]  \right. \nonumber
\\
&& \left. +  \vec U_{\alpha,p}\ofx \vec V_{\beta,p'}^*(\x')\left[\hat
b_p\ofx,\hat b_{p'}^\dagger(\x') \right] +\vec V_{\alpha,p}^*\ofx
\vec U_{\beta,p'}(\x')\left[\hat b_p^\dagger\ofx,\hat b_{p'}(\x')
\right]   \right\}
\end{eqnarray}
gives the condition
\begin{equation}
\sum_p\left( \vec U_{\alpha,p} \vec V_{\beta,p}^*- \vec
V_{\alpha,p}^* \vec U_{\beta,p}\right) = 0.
\end{equation}

\bibliographystyle{unsrt}
\bibliography{Bibliography}
\end{document}